\documentclass[openright,twoside,11pt]{report}

\usepackage[USletter]{vmargin}
\usepackage[latin1]{inputenc}           % Latin1 charset
\usepackage{amsmath}
\usepackage{amssymb}                    % AMS mathematical symbols
\usepackage{graphicx}                   % \includegraphics
\usepackage{fancyhdr}                   % Fancy headers
\usepackage{braket}                     % Bra-ket <0|, |1>
\usepackage{makeidx}                    % Index-generering
\usepackage{tocbibind}                  % Bibl and index in TOC
\makeindex

\setmarginsrb{38mm}{20mm}{25mm}{20mm}{14pt}{11mm}{0pt}{11mm}

\newcommand{\captionfonts}{\small}
\makeatletter  % Makes LaTeX handle @ as a normal letter
\long\def\@makecaption#1#2{%
  \vskip\abovecaptionskip
  \sbox\@tempboxa{{\captionfonts #1: #2}}%
  \ifdim \wd\@tempboxa >\hsize
    {\captionfonts #1: #2\par}
  \else
    \hbox to\hsize{\hfil\box\@tempboxa\hfil}%
  \fi
  \vskip\belowcaptionskip}
\makeatother   % Turn off @ as a normal letter

\newcommand{\im}{\mathrm{i}}
\newcommand{\e}{\mathrm{e}}

\newcommand{\newconcept}[1]{\emph{#1}}

\newcommand{\unity}{\mathbb{I}}
\newcommand{\etal}{{\it et al.~}}

\DeclareMathOperator{\tr}{tr}

\begin{document}

% Indexed synonyms
\index{creation, entanglement of|see{entanglement, of formation}}
\index{density matrix|see{density operator}}
\index{distillable entanglement|see{entanglement, distillable}}
\index{EPR pair|see{Bell state}}
\index{entanglement!of creation|see{entanglement, of formation}}
\index{entanglement!of distillation|see{entanglement, distillable}}
\index{formation, entanglement of|see{entanglement, of formation}}
\index{Local Operations and Classical Communication|see{LOCC}}
\index{LQCC|see{LOCC}}
\index{QLCC|see{LOCC}}

\pagenumbering{roman}
\setcounter{page}{3}
\begin{center}
\thispagestyle{empty}
{\bf Norges teknisk-naturvitenskapelige universitet - NTNU\\
  Fakultet for naturvitenskap og teknologi\\
  Institutt for fysikk}
\vspace{0.5cm}
\par
\resizebox{22mm}{!}{\includegraphics{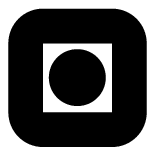}}
\par
\vspace{1cm}
{\LARGE\bf HOVEDOPPGAVE
  \par\vspace{0.2cm}
  FOR}
  \par\vspace{0.8cm}
{\Large\bf STUD. TECHN. GEIR OVE MYHR}\\
\end{center}
\vspace{0.8cm}
\begin{flushleft}
Oppgaven gitt: \hspace{1.5cm}01.02.2004\\
Besvarelsen levert: \hspace{0.9cm}27.06.2004\\
\vspace{1cm}
{\large\bf FAGOMRÅDE: }\hspace{1cm} {\large\bf TEORETISK FYSIKK}\\
\vspace{1cm}
{\large
  Norsk tittel: \hspace{1.5cm}{\it``Mål på sammenfiltring i kvantemekanikk''}\\
}
\vspace{0.3cm}
{\large
  Engelsk tittel: \hspace{1.0cm} {\it``Measures of entanglement in quantum mechanics''}\\
}
\vspace{1cm}
Hovedoppgaven er utført ved Institutt for fysikk, Fakultet for
naturvitenskap og teknologi, under veiledning av professor Jan Myrheim
og stipendiat Stein Olav Skr{\o}vseth.
\end{flushleft}
%\begin{center}\rule{62mm}{0.5pt}\end{center}
\vspace{1.0cm}
\begin{center}
  Trondheim, 27. juni 2004,\\
  \vspace{1.7cm}
  Jan Myrheim\\
  ansvarlig faglærer\\
  professor ved Institutt for fysikk
\end{center}

% Following ISO-7144-1986 the abstract is before the preface
\cleardoublepage
% The \setcounter works around the fact that the abstract environment
% messes up the page numbers.
%\chapter*{Abstract}
\begin{abstract}
  \setcounter{page}{5}
  \thispagestyle{plain} % I want a page number on the abstract page
  \addcontentsline{toc}{chapter}{Abstract}
  I give an overview of some of the most used measures of
  entanglement. To make the presentation self-contained, a number of
  concepts from quantum information theory are first explained. Then the
  structure of bipartite entanglement is studied qualitatively, before a
  number of bipartite entanglement measures are described, both for pure
  and mixed states. Results from the study of multipartite systems and
  continuous variable systems are briefly discussed. 
\end{abstract}

\cleardoublepage
\chapter*{Preface}
\addcontentsline{toc}{chapter}{Preface}

``What's the name again of this quantum effect when one
particle on one side of the universe magically affects a particle on
the other side of the universe?'', I asked while
walking out the door to the corridor. I was on my way to a group of
teenagers who were waiting to hear how it is to study physics at the
university. In case someone asked why I found physics so exciting I could
always mention this quantum, non-locality, spooky
action-at-a-distance thing. If I only remembered the name of it\ldots
``That'd be \emph{entanglement}'', a graduate student next-door suggested,
using the English term. Yes, that was the word lost in the back of my
mind. I had actually never heard a Norwegian name for it.

This happened one of the last days of October 2003. About a month
later it was decided that my master's thesis would be about this very
action-at-a-distance thing, or more precisely, about how to quantify
entanglement. From hardly remembering the name of the game to
understanding how an appropriate measure needs to behave has been a long
and interesting journey. In writing this text I have tried to keep
in mind my own state of knowledge as of
October 2003. With that I hope that any 5th year physics
student will find it sufficiently understandable and self-contained as
to be able to grasp the main points.
Whether I have succeeded or not is for others to decide.

This thesis is the final work leading up to the degree of 
Master of Science in Applied Physics and Mathematics
at the Norwegian University of Science and
Technology (NTNU). It was written during the 2004 spring term. 

I would like to thank Jan Myrheim and Stein Olav Skrøvseth for taking
the extra effort to find a topic in quantum information theory for
my thesis, and for supervising it. I would also like to thank Viktor
H. Havik and Henrik Tollefsen for interesting discussions about
entanglement.

\cleardoublepage
\tableofcontents

\cleardoublepage

\pagestyle{fancy}
%\makeatletter
%\renewcommand{\chaptermark}[1]{\markboth{\@chapapp \ \thechapter:\ #1}{}}
%\makeatother
%\renewcommand{\sectionmark}[1]{\markright{\thesection.\ #1}}
\pagenumbering{arabic}
\setcounter{page}{1}

\chapter{Introduction}
\label{ch:intro}

Quantum entanglement was first viewed as a curiosity when it was
pointed out in the attempt by Einstein, Podolsky and Rosen \cite{EPR:1935} to
show that quantum mechanics could not be a complete theory.
When John Bell many years later published his now
well-known theorem \cite{bell:1964}, it was hardly
noticed in the scientific community. Bell had showed that for a theory
based on local hidden variables, certain correlations were upper
bounded, a result today known as the Bell inequalities. Quantum
mechanics predicted that the inequalities would be violated, thus
giving rise to a way of testing whether the predictions of quantum
mechanics or the assumptions of Einstein \etal were
correct. Experiments have later agreed with quantum mechanics
\cite{aspect:1982},
although there are still critics arguing that there are loopholes in
the experimental assumptions \cite{chaotic_ball,fair_sampling}.

Entanglement has since been regarded as a real -- albeit strange --
phenomenon of quantum mechanics. It was widely regarded as being the
same as violation of some Bell inequality. However, Werner
\cite{werner:1989} showed that there exist mixed quantum
states that allow for a local hidden-variable theory, but
nevertheless are entangled. This led to new criteria to decide if a
state was separable \cite{PPT:1996,HHH:1996}.

Parallel to this, the view of entanglement had gradually changed. From
being regarded as a curiosity, it was now seen as an information processing
and communication resource that could be used for performing tasks
that would be impossible without it. Among the possibilities
introduced were quantum cryptography \cite{qcrypto},
dense coding \cite{dense_coding},
teleportation of
a quantum state \cite{teleportation} and
exponential speed-up of certain computational tasks
\cite{shor_proceedings,shor_journal,grover_proceedings,grover_journal}.
Because entanglement was now viewed as a resource,
it was natural to try to \emph{quantify} the entanglement in quantum
states. 

After the first papers on the quantification of entanglement
\cite{schlienz:1995,bennett1:1996,bennett2:1996,bennett3:1996},
the subject has grown into a whole field of research. The aim of this
work is to describe the entanglement measures that have come out of
this. The main focus is on entanglement between two parties with finite
dimensional Hilbert spaces, as this is the situation where most of the
theory is known. Extensions to entanglement between more than two
parties and entanglement with continuous variables are also briefly
discussed. 

In the course of writing this, several reviews of entanglement and
quantum information theory have been of great help. The textbook by
Nielsen and Chuang \cite{NC00a} gives a nice introduction to the
prerequisites to understand entanglement theory, but contains little
on entanglement itself. John Preskill's lecture notes \cite{preskill}
explain much of the same, but has a whole chapter devoted to entanglement.
The tutorial by Bru{\ss} \cite{bruss-tutorial:2002}
gives a short introduction to most aspects of entanglement, including
some entanglement measures. Both the lecture notes by Eisert
\cite{eisert-lectures:2003} and the review article by Keyl
\cite{keyl:2001} include a chapter on entanglement measures, but the
latter is rather technical in nature. The review article by
M. Horodecki about entanglement measures \cite{mhorodecki:2001}
summarises the most of the field in an elegant way, but it requires much
prior knowledge from the reader, and treats almost exclusively
two-party entanglement.

The rest of this work is organised as follows. Chapter \ref{ch:prelim}
gives an introduction to the concepts needed in the following
chapters. A reader with a basic knowledge in quantum information
theory may skip large portions of this chapter. Chapter \ref{ch:char}
describes two-party entanglement qualitatively, preparing the ground
for chapter \ref{ch:measures} which describes the ways to quantify
two-party entanglement. Finally, in chapter \ref{ch:multipartite} it is
briefly indicated how the notions from the previous chapters may be
extended to systems of continuous variables and to multi-party
systems.

Unless stated otherwise all logarithms are taken base-2.

\chapter{Preliminaries}
\label{ch:prelim}

Before we can understand how to quantify entanglement, we need to have 
some concepts and procedures clear. In this chapter we will have a first look 
at qubits and entangled states, get used to density operators and see how quantum states 
can be transformed into others. 

\section{The qubit}

The simplest nontrivial quantum mechanical system is a two-state system 
which can be described by a vector in two-dimensional complex Hilbert 
space. Such a system is called a \newconcept{qubit}\index{qubit}. 
The favourite qubit of many physicists is the spin 
of a spin-1/2 particle, for instance an electron. 
We will use spins of such particles to illustrate entanglement.

If we measure the $z$-component of the spin, the result is
either up or down. This measurement corresponds to the observable operator
which we call $Z$. After the measurement the state of the particle is an 
eigenstate of the observable. 
We denote these states by $\ket{\uparrow_z}$ and $\ket{\downarrow_z}$, for 
spin parallel to and antiparallel to the $z$-axis. They are mutually orthogonal, 
and we take them as the unit basis vectors of the spin Hilbert space.
In addition to the eigenstates, the particle may be in any 
superposition of the eigenstates, 
\begin{equation}
\label{eq:qubit}
\ket{\psi} = \alpha \ket{\uparrow_z} + \beta \ket{\downarrow_z}
\end{equation}
such that $|\alpha|^2 + |\beta|^2 = 1$. 
When the spin along the $z$-axis is measured, we obtain $+1$ with a probability of 
$|\alpha|^2$ and $-1$ with probability $|\beta|^2$ (where the spin is in 
units of $\hbar$/2). After the measurement the state is collapses into
the corresponding eigenstate.

\begin{figure}[htbp]
  \centering
  \resizebox{0.8\textwidth}{!}{\includegraphics{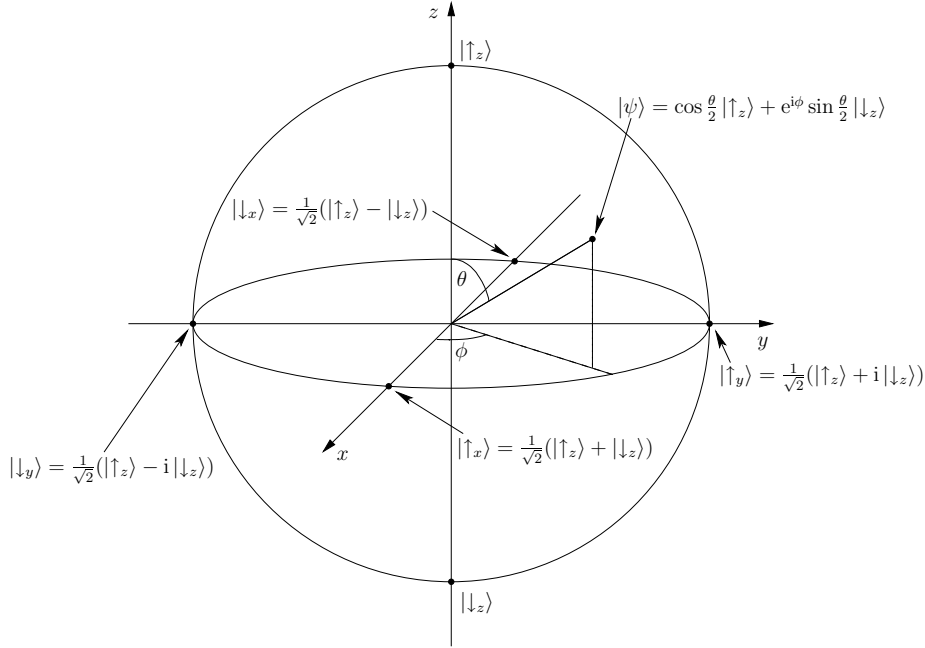}}
  \caption{The Bloch sphere of the Hilbert space spanned by $\ket{\uparrow_z}$ and $\ket{\downarrow_z}$.}
  \label{fig:bloch}
\end{figure}

As the overall phase of a quantum state does not have any physical
significance, only the phase difference between $\alpha$ and 
$\beta$ is important, and for convenience we often choose $\alpha$ to be real. 
In order to also include the property $|\alpha|^2 + |\beta|^2 = 1$, we can 
write the spin state \eqref{eq:qubit} of the particle as
\begin{equation}
\label{eq:qubit_bloch}
\ket{\psi} = \cos\frac{\theta}{2} \ket{\uparrow_z}
+ \e^{\im \phi} \sin\frac{\theta}{2} \ket{\downarrow_z}
.
\end{equation}
This gives a nice and intuitive picture of the state space of the spin-1/2 particle, 
and thereby any qubit. All possible states are covered when the parameters are 
restricted to $0 \leq \phi \leq 2\pi$ and $0 \leq \theta \leq \pi$. Thus, any state 
corresponds to a point on the unit sphere, with the azimuthal angle given by $\phi$ and the 
polar angle given by $\theta$. Such a sphere is called a 
\newconcept{Bloch sphere}\index{Bloch sphere}, and is a very useful tool for visualising 
states of a qubit. A Bloch sphere with some states indicated is shown in figure 
\ref{fig:bloch}. Note that for $\theta = \pi$, the phase of \eqref{eq:qubit_bloch}
changes as $\phi$ changes, but the state is still the same. This is reflected on the 
Bloch sphere, as the azimuthal angle is irrelevant at the poles.

% x-axis, Pauli-matrices
We may choose another set of orthonormal basis vectors. For instance
\begin{equation}
\ket{\uparrow_x} = \frac{1}{\sqrt{2}}\left( \ket{\uparrow_z} + \ket{\downarrow_z}  \right)
\qquad
\ket{\downarrow_x} = \frac{1}{\sqrt{2}}\left( \ket{\uparrow_z} - \ket{\downarrow_z}  \right)
.
\end{equation}
These are the eigenvectors of the $X$-operator, and they are the resulting states if 
we measure the spin along the $x$-axis. Likewise, we may measure the spin along any axis, 
and find a corresponding set of eigenvectors. If we denote the basis vectors
\begin{equation}
\ket{\uparrow_z} = \begin{pmatrix} 1\\0 \end{pmatrix} 
\qquad
\ket{\uparrow_z} = \begin{pmatrix} 0\\1 \end{pmatrix} 
,
\end{equation}
then the operators for the spin along the $x$, $y$ and $z$ axes are nothing
more than the Pauli matrices
\index{X operator}\index{Y operator}\index{Z operator}
\begin{equation}
  \label{pauli_spin}
X \equiv \begin{pmatrix} 0 & 1 \\ 1 & 0 \end{pmatrix}
\qquad
Y \equiv \begin{pmatrix} 0 & -\im \\ \im & 0 \end{pmatrix}
\qquad
Z \equiv \begin{pmatrix} 1 & 0 \\ 0 & -1 \end{pmatrix}
.
\end{equation}
For the spin along any other unit vector, say 
$a\vec{x} + b\vec{y} + c\vec{z}$ where $a^2 + b^2 + c^2 = 1$, the 
spin operator is $aX + bY + cZ$. All operators of this form have eigenvalues 
$\pm 1$ corresponding to spin parallel and antiparallel to the vector.

% General qubit
As already mentioned, other systems are equally good qubits as a spin-1/2 particle. 
Among the more popular ones are the polarisation of a photon and the energy 
levels of a two-level system. They can be represented by a Bloch sphere just as well
as the spin of an electron can. The advantage of thinking about a spin-1/2 particle is 
that the vectors on the Bloch sphere corresponds to actual spin directions in real 
space. Thus, a qubit in a state close to the $y$-axis on the Bloch sphere, has, when the
qubit is a spin-1/2 particle, a spin pointing in a direction close to the $y$-axis.

In general the particular implementation of the qubit is not
important for our purpose, so we simply
denote the basis vectors $\ket{0}$ and $\ket{1}$, which in the case of a spin-1/2 
particle are typically $\ket{\uparrow_z}$ and $\ket{\downarrow_z}$, respectively. The 
superpositions of these states corresponding to the spin along the $x$-axis, are
often denoted $\ket{+}$ and $\ket{-}$, according to the sign in the superposition.
\begin{equation}
\ket{+} = \frac{1}{\sqrt{2}}\left( \ket{0} + \ket{1} \right)
\qquad
\ket{-} = \frac{1}{\sqrt{2}}\left( \ket{0} - \ket{1} \right)
\end{equation}

Three qubit operators are of special importance, namely the
unitary 
Pauli operators \index{Pauli operators}
of \eqref{pauli_spin}. In the context of
general qubits, they can be written as
\index{X operator}\index{Y operator}\index{Z operator}
\begin{subequations}
  \begin{align}
    \label{eq:pauli}
    X &\equiv \ket{1}\bra{0} + \ket{0}\bra{1},\\
    Y &\equiv \im \ket{1}\bra{0} - \im \ket{0}\bra{1},\\
    Z &\equiv \ket{0}\bra{0} - \ket{1}\bra{1}
    .
  \end{align}
\end{subequations}
$X$ is also called the
\newconcept{bit flip operator}\index{bit flip operator} as it turns a
$\ket{0}$ into a $\ket{1}$ and vice versa. $Z$ is called the
\newconcept{phase flip operator}\index{phase flip operator} as it
switches between the states $\alpha \ket{0} + \beta \ket{1}$ and
$\alpha \ket{0} - \beta \ket{1}$. The Pauli operators are defined
with respect to a particular basis. For instance, in the
$\{\ket{+},\ket{-}\}$ basis, $Z$ takes the role of the bit flip
operator.

The name \emph{qubit} comes from \emph{quantum bit} and it plays the same role
in quantum information theory as the bit does in classical information theory. The 
bit is a system which can be in one of two states, usually denoted 0 and 1. Thus, 
each bit is capable of storing enough information to answer exactly one yes/no 
question. As we have just seen, the qubit is a much richer structure. In addition to 
being capable of being in the states $\ket{0}$ and $\ket{1}$, just as a bit, it can be 
in a continuum of superpositions between those states, given by two parameters. 
Still, only a single bit of information can be read from a single qubit. When the 
spin of a spin-1/2 particle is measured along an axis the only possible results are
parallel and antiparallel to the chosen axis. After the measurement the state is 
changed, and from the result of the measurement we know what it is. No more 
information about the original state can be retrieved.

\section{Entangled states}
\label{sec:prelim_entangled}

From the simplest quantum systems whatsoever, we now turn to look at the
simplest \emph{composite}\index{composite system} quantum systems. A
composite system is a system which consists of two or more parts, and
the simplest one is a system consisting of two qubits. We call the two
systems $A$ and $B$.
In communication and entanglement theory the convention is to think of
the subsystems as being in the possession of two well separated
observers, called Alice and Bob\index{Alice and Bob}.
Any state of each of the two systems can be
written as
\begin{equation}
  \label{eq:twoqubits}
  \ket{\psi}_A = \alpha \ket{0}_A + \beta \ket{1}_A
  \qquad
  \ket{\phi}_B = \gamma \ket{0}_B + \delta \ket{1}_B
\end{equation}
with $|\alpha|^2 + |\beta|^2 = 1$ and $|\gamma|^2 + |\delta|^2 = 1$
The composite state of the two systems is then simply the tensor
product (or direct product) of the two states.
\begin{equation}
  \label{eq:twoqubitscomposite}
  \ket{\Psi_\text{prod}} = \ket{\psi}_A \otimes \ket{\phi}_B
\end{equation}
Such a state is called a
\newconcept{product state}\index{product state},
but product states are not the
only physically realisable states. If we let the two systems interact
with each other, any superposition of product states is realisable.
Hence, a general composite state can be written as
\begin{equation}
  \label{eq:entangled_state}
  \ket{\Psi} = \sum_{ij} \alpha_{ij} \ket{\psi_i}_A \otimes \ket{\phi_j}_B
\end{equation}
where $\sum |\alpha_{ij}|^2 = 1$ and the sets
$\{\ket{\psi_i}\}$ and $\{\ket{\phi_j}\}$ are orthonormal bases for
the two subsystems.
Any composite state that is not a
product state is called an
\newconcept{entangled state}\index{entangled state}.

A composite quantum state consisting of two parts only, is called a
\newconcept{bipartite state}\index{bipartite state}, as opposed to
\newconcept{multipartite states}\index{multipartite state} which
consist of more than two parts. For bipartite qubit states, four
entangled states play a major role, namely the singlet state
\begin{subequations}
\begin{equation}
  \label{eq:bell-singlet}
  \ket{\Psi^-} \equiv \frac{1}{\sqrt{2}}(\ket{01}-\ket{10})
\end{equation}
and the three triplet states
\begin{align}
  \label{eq:label-triplet}
  \ket{\Psi^+} &\equiv \frac{1}{\sqrt{2}}(\ket{01}+\ket{10})\\
  \ket{\Phi^-} &\equiv \frac{1}{\sqrt{2}}(\ket{00}-\ket{11})\\
  \ket{\Phi^+} &\equiv \frac{1}{\sqrt{2}}(\ket{00}+\ket{11})
\end{align}
\end{subequations}
where we have used $\ket{ij}$ as a shorthand notation for $\ket{i}
\otimes \ket{j}$.
They are called \newconcept{Bell states}\index{Bell states} or
\newconcept{EPR pairs}. Together they form an orthonormal basis
for the state space of two qubits, called
the \newconcept{Bell basis}\index{Bell basis}. The Bell states are
maximally entangled and one can be converted into another by applying
a unitary transform locally on any one of the subsystems.

Note that if we measure the state of one qubit in a Bell state (that
is, measure the $Z$ operator which has eigenvalues $\pm 1$), we
immediately know the state of the other particle. In the singlet
Bell state, a measurement of qubit A will yield one of the eigenstates
$\ket{0}$ and $\ket{1}$, each with a probability of 1/2. These
results leave qubit B in state $\ket{1}$ or $\ket{0}$, respectively.

For a single qubit we could always change to another basis
where the outcome of a $Z$ measurement would be
given. For a spin-1/2 particle this means that the spin is always
pointing in \emph{some} direction, even though the state will show up as a
superposition in a basis where the state is not one of the basis
states. If the particle is entangled with another particle, though,
the direction of the spin of that particle alone is not well
defined. Actually, for particles in one of the Bell states, the probability
for measuring the spin of the particle to ``up'' (while ignoring the
other particle) is 1/2 for \emph{any} direction. 

\section{Density operators and mixed states}
\label{sec:mixed}

We want to be able to describe the state of a subsystem that is entangled with
another subsystem to which we do not have access.
State vectors cannot be used for this purpose, and we need another
representation of quantum states,
namely \newconcept{density operators}\index{density operator}.
This is not only useful for describing ``locally'' a subsystem which
is part of an entangled system for which we know the state. The
formalism is also necessary for describing quantum mechanical
experiments -- where noise is inevitable --
and doing quantum statistical mechanics. In
the laboratory it is impossible to isolate the quantum systems under
study completely. The systems become entangled with the environment
through unwanted, but unavoidable, interactions. Nevertheless, it is
necessary to describe the system without taking the environment into
account.

In the density operator formalism, we describe quantum states by
operators on the system's Hilbert space instead of unit vectors on
it. For any quantum state vector $\ket{\psi}$, the corresponding
density operator is the projection operator $\ket{\psi}\bra{\psi}$. 
As linear operators may be -- and often is -- represented by matrices, density
operators are also called
\newconcept{density matrices}\index{density matrix}.

So far we have only represented the state in a new way, using density
operators instead of vectors. But what if we do not know exactly in
what state our system is? Imagine, for example, that we have a
machine in our lab which outputs particles in state $\ket{\psi}$ (think
of it as some component of a spin). But once in a while events beyond
our control and knowledge (a spike in the electric grid, the turning
on of an electromagnet in a neighbouring lab, etc.) make the machine
malfunction slightly producing states which are close to, but not
exactly $\ket{\psi}$.
Let us for simplicity say that the state
$\ket{\psi^\prime}$ is
always created when the machine malfunctions, and that the
probability for that to happen to any given particle is $p$. Given
this uncertainty, the state cannot be described by a state
vector. Still, in terms of density matrices it is described as
\begin{equation*}
  \rho = (1-p)\ket{\psi}\bra{\psi} + p \ket{\psi^\prime}\bra{\psi^\prime}
       .
\end{equation*}

A state that may be represented by a unit vector is called a
\newconcept{pure state}\index{pure state}. For a pure state we have
\newconcept{maximal knowledge}\index{maximal knowledge} of the
state. Other states are \newconcept{convex combinations}\footnote{
  A convex combination of elements $x_1, x_2, \ldots, x_N$ is an
  element which can be written $\sum_i p_i x_i$ for $p_i \geq 0$ and
  $\sum_i p_i = 1$. A convex combination of two points is a point on
  the straight line connecting them. 
}
of pure states, and are called
\newconcept{mixed states}\index{mixed state}.
To check if a given density operator, $\rho$, represents a pure state, it is
sufficient to check if $\rho = \rho^2$, which holds for all pure
states and no mixed states. 

The density operator does not tell us
from what pure states the state was prepared.
For example, we can have a state $\ket{0}$ or $\ket{1}$,
each with probability 1/2. The density operator is then
$\frac{1}{2} (\ket{0}\bra{0} + \ket{1}\bra{1}) = \frac{1}{2} \unity$.
The same density operator can be made by mixing the states
$\ket{+} = \frac{1}{\sqrt{2}} (\ket{0} + \ket{1})$ and
$\ket{-} = \frac{1}{\sqrt{2}} (\ket{0} - \ket{1})$ in equal amounts, or 
the three states $\ket{0}$, $\ket{+}$ and
$\frac{1}{2}\ket{0} - \frac{\sqrt{3}}{2}\ket{1}$ in amounts $p_1 =
\frac{1}{2}-\frac{\sqrt{3}}{6}$, $p_2 = \frac{\sqrt{3}}{3+\sqrt{3}}$ and
$p_3 = 1- \frac{\sqrt{3}}{3}$, respectively.
We call these different
collections of pure states with corresponding probabilities,
$\{p_i,\ket{\psi_i}\}$,
\newconcept{ensembles}\index{ensemble}. Sometimes it is also called a
\newconcept{realisation}\index{realisation}\index{density operator!realisation of}
and we say that an ensemble
\emph{realises} a mixed state. Hence, when we describe a state from an
ensemble by its density operator, we discard the information about
which ensemble the mixed state was made from. The density operator
still describes the mixed state as well as can be done, as states from
different ensembles having the same density operator are experimentally
indistinguishable.

%% Nytt forsøk
%We call these different collections of pure states with corresponding
%probabilities, $\{p_i,\ket{\psi_i}\}$, pure state
%\newconcept{realisations}
%\index{realisation}\index{density operator!realisation of}
%of a density operator. In general, a realisation of the mixed state
%$\rho$ is a collection of
%(generally mixed) states and corresponding probabilities,
%$\{p_i,\rho_i \}$, such that $\rho = \sum_i p_i \rho_i$.
%A concept closely related to (and often not distinguished from) a
%realisation is an
%\newconcept{ensemble}\index{ensemble}. An ensemble is also a
%collection of probabilities and states
%$\{p_i,\rho_i \}$, and can be thought of as a
%large collection of systems in different states, and if one is chosen
%at random, the probability for it to be in state $\rho_i$ is $p_i$. 
%What separates it from a mixed state is that each system carries a
%label telling what state it is in, so even though the state is unknown
%{\it a priori}, we can perform different operations on different
%states. This can be illustrated by two machines (sources) spitting out
%a particle in a quantum state each second. The probability for them to
%emit a particle in state $\rho_i$ is $p_i$. The only difference
%between the machines is that one of them has a display indicating
%which state it is emitting, whereas we cannot tell what state comes
%out of the other one. The first machine, then, emits an ensemble and
%the second mixed states.
%% Avslutning som på forrige avsnitt

This indistinguishability makes the density operator a perhaps more
intuitive representation of a state than the state vector. 
For example, a state vector has
an arbitrary overall phase factor, so two vectors differing by a phase factor
represent the same state. Still, when the vector is put together
with its dual vector to form a pure state density operator, the
phase factor vanishes because of the complex
conjugation. Hence, all state vectors
representing the same state are represented by the same density
operator. 

To complete our discussion about density operators, we should take
note of some of their properties. For a more thorough discussion, see
e.g.~\cite{NC00a}.
Density operators are (i)
\newconcept{positive}\index{positive operator}\footnote{
  Positive operators here refer to \emph{positive semidefinite}
  operators. \emph{Positive definite} operators, on the other hand, satisfy
  $\bra{v}\rho\ket{v} > 0$ for any $\ket{v}$ and have only positive
  eigenvalues.
}, meaning that for any vector $\ket{v}$,
$\bra{v}\rho\ket{v} \geq 0$, or equivalently that it is Hermitian with
nonnegative eigenvalues.
Also, (ii) the trace of a density operator is
unity, $\tr(\rho) = 1$. It turns out that any operator satisfying
these two criteria can be realised by a pure state ensemble.
Hence, we may take
these criteria as the defining properties of a density operator.

We will now see how the density operators are used to describe
individual parts of an entangled system. In fact, we will see that
even when the composite system is in a pure state, if the subsystems
are entangled, the individual subsystems are in mixed states.

Consider a bipartite composite system of two qubits, shared
between Alice and Bob. The qubits are entangled, and the system is in
the state
\begin{equation}
  \label{eq:example_partial_state}
  \ket{\psi} = \alpha\ket{00} + \beta\ket{11}
  .
\end{equation}
Now, imagine that Alice loses contact with Bob, and wants to describe
her system without any reference to the qubit in Bob's lab.
Whatever Bob happens to be doing to his qubit should not have
observable consequences in Alice's lab. If it had, it could have been
used to perform superluminal communication in some reference frame
which would wreak havoc in
physics by allowing consequences to happen before the cause\footnote{
  This is of course in itself no compelling reason. Even John Bell
  was afraid that entanglement might do away with
  relativity \cite{whitaker:1998}. The real reason is that quantum
  mechanics tells us that the probabilities for outcomes of local
  measurements on one system do not change when an arbitrary
  operation is performed on the other system.
}.
One of the things Bob might choose to do with his system, is to measure
$Z$ in the $\{\ket{0},\ket{1}\}$ basis. If he does, he will get
$\ket{0}$ or $\ket{1}$ 
with probability $|\alpha|^2$ or $|\beta|^2$, respectively. This
leaves Alice's qubit in the same state as the one Bob measured, but as
she wouldn't know the result, the state of her qubit would be a
mixture of $\ket{0}$ and $\ket{1}$, namely
\begin{equation}
  \label{eq:example_partial_A}
  \rho_A = |\alpha|^2 \ket{0}\bra{0} + |\beta|^2 \ket{1}\bra{1}
\end{equation}
This mixture correctly predicts the probabilities for all
measurements that can be performed locally by Alice, regardless of
what Bob really does to his system.

Formally, ignoring some degrees of freedom in a system
(like the ones corresponding to a particle out of reach)
is done by
\newconcept{tracing out}\index{trace out} the relevant degrees of
freedom from the density operator. This is also called taking the
\newconcept{partial trace}\index{partial trace} with respect to the
degrees of freedom that are to be traced out.
Any bipartite state can be written as
\begin{equation*}
  \rho =
  \sum_{ijkl} c_{ijkl} \ket{a_i}\bra{a_j} \otimes \ket{b_k}\bra{b_l}
\end{equation*}
and the partial trace with respect to
system $B$ of a bipartite system is defined as
\begin{align}
  \label{eq:partial_trace}
  \tr_B(\rho)
  &\equiv \sum_{ijkl} c_{ijkl} \ket{a_i}\bra{a_j}
  \otimes \tr\big(\ket{b_k}\bra{b_l}\big) \notag \\
  &= \sum_{ijkl} c_{ijkl} \braket{b_l|b_k} \ket{a_i}\bra{a_j} \notag\\
  &= \sum_{ij} C_{ij} \ket{a_i}\bra{a_j}
\end{align}
with $C_{ij} = \sum_{kl} c_{ijkl} \braket{b_l|b_k}$.
The density operator obtained by tracing out one part of the system is
called the
\newconcept{reduced density operator}\index{density operator!reduced}. 

We can perform the partial trace on our example state
\eqref{eq:example_partial_state}. The density matrix is
\begin{equation}
  \label{eq:example_partial_matrix}
  \rho = |\alpha|^2 \ket{00}\bra{00} +
  \alpha \beta^* \ket{00}\bra{11} +
  \alpha^* \beta \ket{11}\bra{00} +
  |\beta|^2 \ket{11}\bra{11}
  ,
\end{equation}
so the reduced density matrix becomes
\begin{align}
  \rho_A
  &= \tr_B(\rho)\notag \\
  &= |\alpha|^2 \braket{0|0} \ket{0}\bra{0} +
  \alpha \beta^* \braket{0|1} \ket{0}\bra{1} +
  \alpha^* \beta \braket{1|0} \ket{1}\bra{0} +
  |\beta|^2\braket{1|1} \ket{1}\bra{1} \notag \\
  &= |\alpha|^2 \ket{0}\bra{0} + |\beta|^2\ket{1}\bra{1},
\end{align}
which is just what we expected.

% Which mixed states are entangled
In section \ref{sec:prelim_entangled} we saw what we meant by an
entangled state as long as the states were pure. But this leaves the
question of when a mixed state is entangled. Consider for instance the
mixture of two Bell states
\begin{equation}
  \label{eq:bell_mixture}
  \rho = \frac{1}{2} \ket{\Phi^+}\bra{\Phi^+} +
  \frac{1}{2} \ket{\Phi^-}\bra{\Phi^-}
  .
\end{equation}
Written out in an orthonormal product basis, this is
\begin{align}
  \rho
  &=
  \begin{aligned}[t]
    &\frac{1}{4} (\ket{00}\bra{00} + \ket{00}\bra{11}
    + \ket{11}\bra{00} + \ket{11}\bra{11}) \notag \\
    &+\frac{1}{4} (\ket{00}\bra{00} - \ket{00}\bra{11}
    - \ket{11}\bra{00} + \ket{11}\bra{11})\notag
  \end{aligned}\\
  &= \frac{1}{2} \ket{00}\bra{00} + \frac{1}{2} \ket{11}\bra{11}
\end{align}
which is a mixture of the product states $\ket{00}$ and
$\ket{11}$. So the mixed state can be realised by both
an ensemble of maximally entangled states and an ensemble of 
product states. We say that a mixed state is
\newconcept{separable}\index{separable state} if and only if it can be realised
as a mixture (convex combination) of locally prepared states\footnote{
  This definition is due to Werner \cite{werner:1989} who called them
  ``classically correlated'' states. 
}. That is, it can be written as\footnote{
  In fact it was considered sufficient that that such a sum
  \emph{approximated} the state arbitrarily well. It was shown in
  \cite{HHH:1996}, though, that in the finite dimensional case any
  separable state could be expressed in $d^2$ terms, where $d$ is the
  dimension of the Hilbert space of the composite system.
}
\begin{equation}
  \label{eq:def_separable}
  \sum_i p_i \, \rho_i^{(A)} \otimes \rho_i^{(B)}
\end{equation}
where $\{p_i\}$ forms a probability distribution; $p_i \geq 0$ and
$\sum_i p_i = 1$.
This also means (\cite{HHH:1996}) that the state can be written as a
mixture of pure product states,
\begin{equation}
  \label{eq:product_separable}
  \sum_{kl} p_{kl} \ket{\psi_k^{(A)}} \bra{\psi_k^{(A)}}
  \otimes \ket{\psi_l^{(B)}} \bra{\psi_l^{(B)}}
  .
\end{equation}
Hence, a mixture of entangled states need not be entangled, but a
mixture of separable states is always separable. Mixture is a process
which destroys entanglement. This is because by discarding information
about which of a number of entangled states the system is in, it can no
longer be distinguished from a mixture of separable states.

\section{The Schmidt decomposition}
\label{sec:schmidt}

A useful tool when working with bipartite, pure state entanglement is
the
\newconcept{Schmidt decomposition}\index{Schmidt decomposition}.
It is a decomposition into the biorthogonal basis which gives the
smallest possible number of terms for a product basis.

Given a bipartite pure state $\ket{\psi}$, we may write it in terms of
some product basis, orthonormal in both subsystems
\begin{equation}
  \label{eq:schmidt_general_form}
  \ket{\psi} = \sum_{i=1}^{d_A} \sum_{j=1}^{d_B} c_{ij} \ket{a_i} \otimes \ket{b_j}
  .
\end{equation}
The coefficients $\{c_{ij}\}$ may be seen as elements in a
$d_A \times d_B$ matrix $C$.
Using the singular value decomposition this may be
written as
\begin{equation}
  \label{eq:schmidt_svd}
  C = UDV
\end{equation}
where $D$ is a $d_A \times d_B$ matrix which is zero except for the
diagonal elements which are real and positive, $U$ is a $d_A \times d_A$
unitary matrix and $V$ a $d_B \times d_B$ unitary matrix. This means
that $c_{ij}$ may be written as
$\sum_{k=1}^{\min(d_A,d_B)} u_{ik} d_{kk} v_{kj}$, so the state vector
may be written
\begin{align*}
  \ket{\psi}
  &= \sum_{i=1}^{d_A} \sum_{j=1}^{d_B} \sum_{k=1}^{\min(d_A,d_B)}
  u_{ik} d_{kk} v_{kj} \ket{a_i} \otimes \ket{b_j}\\
  &= \sum_{k=1}^{\min(d_A,d_B)}
  d_{kk}
  \left(\sum_{i=1}^{d_A} u_{ik} \ket{a_i}\right)
  \otimes
  \left(\sum_{j=1}^{d_B} v_{kj} \ket{b_j}\right)
  .
\end{align*}
We now take as our new basis vectors
\begin{equation*}
  \ket{a_k^\prime} \equiv \left(\sum_{i=1}^{d_A} u_{ik} \ket{a_i}\right)
  \qquad
  \ket{b_k^\prime} \equiv \left(\sum_{j=1}^{d_B} v_{kj} \ket{b_j}\right)
\end{equation*}
and define $\lambda_k = d_{kk}$ to obtain the final form
\begin{equation}
  \label{eq:schmidt_decomposed}
  \ket{\psi} = \sum_{k=1}^{\min(d_A,d_B)}
  \lambda_k \ket{a_k^\prime} \otimes \ket{b_k^\prime}
\end{equation}
which is the Schmidt decomposition. What we have achieved is to write
any state as a linear combination of maximum $\min(d_A,d_B)$ product
vectors. Also, not only are the product vectors in the basis
orthogonal, the basis vectors from each subsystem are only used
once. These vectors
are still orthogonal as we have only performed a unitary
transformation on the set of basis vectors, making the set of new
basis vectors \newconcept{biorthogonal}\index{biorthogonal}.
The coefficients $\{\lambda_k\}$ are called the
\newconcept{Schmidt coefficients}\index{Schmidt coefficients}, although
in the literature this name is sometimes given to $\{\lambda_k^2\}$

For many purposes the mere existence of the Schmidt decomposition is
enough to make it usable, whereas it sometimes needs to be calculated
explicitly. An example of the former is to show that the eigenvalues
of the two reduced density matrices of a pure, bipartite state are
equal. Actually, it can be seen directly from
\eqref{eq:schmidt_decomposed} that by forming the density matrix and
tracing out \emph{any} one of the two subsystems, the reduced matrix is already on
diagonal form with $\{\lambda_k^2\}$ as eigenvalues. It is often
easier to calculate the Schmidt coefficients this way than doing the
singular value decomposition explicitly.

\section{Generalised measurements}
\index{measurement!generalised|(}

To facilitate the quantum operations we will generalise the
concept of measurements from the way it is usually treated in courses
of introductory quantum mechanics. In that formalism, which is called
\newconcept{projective measurements}\index{measurement!projective},
the objects representing the measurements
are Hermitian operators. The result of the measurement is an
eigenvalue of the observable and after the measurement,
the new state of the system is in the eigenstate corresponding to the
eigenvalue (or projected into the eigenspace if the eigenvalue is
degenerate). The probability for an outcome of the measurement is the
absolute squared expansion coefficient when the state is expanded in
the eigenstates of the observable.

In the general measurement scheme, a measurement is represented by a
set of linear operators $\{M_i\}$
called \newconcept{measurement operators}\index{measurement operators}
satisfying the completeness relation
\begin{equation}
  \label{eq:measurement_completeness}
  \sum_m M_m^\dagger M_m = 1
  .
\end{equation}
The probability for the outcome to be $m$ is
\begin{equation}
  \label{eq:measurement_probability}
  p(m)=\bra{\psi}M_m^\dagger M_m\ket{\psi}
\end{equation}
and after the measurement with result $m$, the new state of the system
is
\begin{equation}
  \label{eq:measurement_newstate}
  \frac
  {M_m\ket{\psi}}
  {\sqrt{\bra{\psi}M_m^\dagger M_m\ket{\psi}}}
  .
\end{equation}
We may, at least in theory, implement any measurement that can be
described by a set of measurement operators. Thus, we may construct
measurements which convert an initial state to a desired state with a
certain probability, discarding the system if that particular outcome
does not occur. As we will see, this is one of the ideas behind
distillation of entanglement.

% Connection to projective measurement
The projective measurement formalism is a special case of the
generalised measurement scheme. The observable, $O$, is a Hermitian
operator and can be decomposed according to the spectral theorem.
\begin{equation}
  \label{eq:obs_decomp}
  O = \sum_m \lambda_m \ket{v_m}\bra{v_m}
\end{equation}
where $\{\ket{v_m}\}$ are the eigenvectors of the observable, and
$\{\lambda_m\}$ the corresponding eigenvalues. By taking as
measurement operators $M_m = \ket{v_m}\bra{v_m}$, the completeness relation
\eqref{eq:measurement_completeness} is satisfied. The probability for a
given outcome, $m$, of a measurement is
\begin{equation*}
  \bra{\psi}M_m^\dagger M_m\ket{\psi}
  = \braket{\psi|v_m} \braket{v_m|v_m} \braket{v_m|\psi}
  = |\braket{v_m|\psi}|^2
\end{equation*}
which is exactly the absolute square of the expansion coefficient of
$\ket{\psi}$ for that eigenstate. The state after the measurement is
\begin{equation*}
  \frac{M_m\ket{\psi}}{\sqrt{\bra{\psi}M_m^\dagger M_m\ket{\psi}}}
  = \frac{\braket{v_m|\psi}\ket{v_m}}{|\braket{v_m|\psi}|}
  = \e^{\im \theta}\ket{v_m}
\end{equation*}
with an arbitrary phase $\theta$. Like $\ket{v_m}$, this is an
eigenstate of $O$.

% Gen. measurements of Density matrices for pure states
The generalised measurement scheme can easily be further generalised
to density operators. For pure states, the generalisation is
straightforward.
The density operator corresponding to the pure state vector $\ket{\psi}$
is $\rho = \ket{\psi}\bra{\psi}$.
The probability for a given outcome is of course still the
same as in \eqref{eq:measurement_probability}, but we
write it in terms of the density operator:
\begin{equation}
  \label{eq:measurement_probability_op}
  p(m)=\tr(M_m\rho M_m^\dagger)
\end{equation}
When the state changes after a
measurement with result $m$, the density operator corresponding to the
state vector \eqref{eq:measurement_newstate} is obviously
\begin{equation}
  \label{eq:measurement_newstate_op}
  \frac{M_m\ket{\psi}\bra{\psi}M_m^\dagger}{\bra{\psi}M_m^\dagger  M_m\ket{\psi}}
  =
  \frac{M_m\rho M_m^\dagger}{\tr(M_m\rho M_m^\dagger)}
\end{equation}

% Gen. measurements of Density matrices for mixed states
The probability \eqref{eq:measurement_probability_op} and new state
\eqref{eq:measurement_newstate_op} is also valid if the density
operator represents a mixed state. To see this, we write the density
operator in terms of one of its pure state realisations:
\begin{equation}
  \label{eq:mixed_realisation}
  \rho = \sum_i p_i \ket{\psi_i}\bra{\psi_i}
  = \sum_i p_i \rho_i
\end{equation}
The joint probability for the system to be in the pure state $\ket{\psi_i}$
and measuring a result $m$ is $p_i p(m|i)$. Summing over all $i$ we
get the total probability for getting $m$ as the result of the
measurement.
\begin{align*}
  p(m)
  &= \sum_i p_i \tr(M_m\rho_i M_m^\dagger)\\
  &= \tr \bigg(M_m \Big[\sum_i p_i \rho_i \Big] M_m^\dagger \bigg)\\
  &= \tr (M_m \rho M_m^\dagger)
\end{align*}
Similarly, the new state after the measurement of the mixed state
becomes
\begin{equation*}
  \rho_{\text{out}}
  = \sum_i p_i M_m \rho_i M_m^\dagger 
  = M_m \Big[\sum_i p_i \rho_i \Big] M_m^\dagger
  = M_m \rho M_m^\dagger
  .
\end{equation*}

% Any operation may be implemented as a gm.
The general measurement scheme covers all operations that can be
performed on a quantum system. A unitary transformation is the special
case of a one-outcome measurement as the measurement operators then
will have to be unitary to satisfy the completeness relation
\eqref{eq:measurement_completeness}. Performing a different unitary
transformation $U_m$ conditional on the outcome of a
measurement described by the operators $\{M_m\}$, is described by the
new measurement operators $\{U_m M_m\}$. It is easy to see that these
operators satisfy the completeness relation, that the probabilities of the
outcomes are the same and that the post measurement state is what we
expect, namely $U_m (M_m \rho M_m^\dagger ) U_m^\dagger$. Similarly,
consecutive measurements can be put together to form a single set of
measurement operators, even when the measurements to be done depend on
the outcome of the previous ones. When doing this, all possible
combinations of
measurement outcomes are considered as single outcomes of the total
measurement. 

In applications it depends on the situation whether we use
the full generalised measurement scheme or only consider projective
measurements. Projective measurements are sometimes simpler to handle
and combined with unitary transformations they describe everything
that can be described by generalised measurements. Therefore, both
formalisms will be used in the following.

\index{measurement!generalised|)}

\section{Transformation of states and LOCC}

We will see later that the set of operations that we allow to
be performed on a quantum state is essential to the
most basic entanglement measures.
All operations can be described as generalised measurements. But we
will find it useful to describe them as measurements combined with
unitary transformations where the next step in a sequence of operations
may be decided by a previous measurement outcome.

\subsection{Local Operations and Classical Communication}
In the context of entanglement a subclass of quantum operations is
particularly important, namely local quantum operations (LO). When we have a
bipartite or multipartite state, the subsystems may be separated, or
at least well isolated from each other. We may therefore only perform
operations that act locally on each subsystem. That is, the operators
will be of the form $O_A \otimes O_B \otimes \cdots$. In addition, the
parties sharing a multipartite state are usually allowed to
communicate classically, and through this classical communication
(CC) they may perform different operations depending on the
measurement results of the other parties. This subclass of operations
is called
\newconcept{Local Operations and Classical Communication}, usually
shortened to \newconcept{LOCC}\index{LOCC}. In the literature it is
sometimes named otherwise, e.g.~QLCC or LQCC.
LOCC is important in entanglement theory because entanglement is
exactly those correlations that cannot be created using LOCC.

\subsection{Transformation of a Bell state}
Suppose Alice and Bob share a Bell state
$\ket{\Phi^+} = \frac{1}{\sqrt{2}} \ket{00} + \frac{1}{\sqrt{2}}
\ket{11}$
and want to transform it into the state
$\cos\theta \ket{00} + \sin \theta \ket{11}$. One way of doing this
is as follows.
Alice performs a measurement described by the measurement operators
\begin{equation}
  \label{eq:measurement_op_ex1}
  M_1 = \begin{pmatrix} \cos\theta & 0 \\ 0 & \sin\theta \end{pmatrix}
  \qquad
  M_2 = \begin{pmatrix} \sin\theta & 0 \\ 0 & \cos\theta \end{pmatrix}
\end{equation}
where the local state vectors are denoted as usual:
\begin{equation}
  \label{eq:ket-vector}
  \ket{0} = \begin{pmatrix} 1 \\ 0 \end{pmatrix}
  \qquad
  \ket{1} = \begin{pmatrix} 0 \\ 1 \end{pmatrix}
  .
\end{equation}
The operators clearly satisfies the completeness relation
\eqref{eq:measurement_completeness}, as
\begin{equation*}
  \begin{pmatrix} \cos^2\theta & 0 \\ 0 & \sin^2\theta \end{pmatrix}
  +
  \begin{pmatrix} \sin^2\theta & 0 \\ 0 & \cos^2\theta \end{pmatrix}
  =
  \begin{pmatrix} 1 & 0 \\ 0 & 1 \end{pmatrix}
  .
\end{equation*}
After the measurement we may have one of two states, depending on the
outcome,
\begin{equation*}
  \ket{\psi_1} = \cos \theta \ket{00} + \sin \theta \ket{11}
  \qquad
  \ket{\psi_2} = \sin \theta \ket{00} + \cos \theta \ket{11}
\end{equation*}
where $\ket{\psi_1}$ is the desired output state. If the result of the
measurement is 2, leaving the system in the state $\ket{\psi_2}$,
Alice applies the $X$ operator (also called
bit flip operator\index{bit flip operator})
to her system
converting $\ket{0}$ into $\ket{1}$ and vice versa. Then she tells Bob
to do the same through the classical communication channel. The
resulting state is $\ket{\psi_1}$ and their common goal is achieved.

The transformation we just did, transforms a maximally entangled state
into a less or equally entangled state. For $\theta = 0$ and
$\theta = \pi/2$ the output state is a product state, which is not
entangled at all, whereas for $\theta = \pi/4$ and $\theta = 3\pi/4$
the output state is also a Bell state. 

\subsection{Concentrating partial pure state entanglement}
\label{sec:concentrating}
We may also try to do the transformation the other way around. We
begin with the state $\ket{\psi_1}$, where we for simplicity suppose
that $0 < \theta < \pi/4$, and want to end up with the Bell state
$\ket{\Phi^+}$.
The lower boundary of this interval is a
product state for $\theta = 0$ which cannot possibly be turned into an
entangled state by LOCC. The upper boundary $\theta = \pi/4$ is the
desired output state and will need no transformation.

The problem is now to construct local measurement operators which give
$\ket{\Phi^+}$ as a post measurement state. The operator
\begin{equation*}
  M_1 = \text{const} \cdot
  \begin{pmatrix}
    \frac{1}{\cos\theta} & 0 \\ 0 & \frac{1}{\sin\theta}
  \end{pmatrix}
\end{equation*}
will give the desired output, and we want the constant to be as large as possible
to maximise the probability for this output. Still, for the
completeness relation \eqref{eq:measurement_completeness} to be satisfied
none of the diagonal elements can be greater than 1. This is because
for any $M_m$, $M_m^\dagger M_m$ has nonnegative diagonal elements,
and each diagonal element needs to sum up to 1 to get the identity
operator in \eqref{eq:measurement_completeness}. In the valid range of
$\theta$, $\cos \theta > \sin \theta$, so the maximal constant is
$\sin \theta$. This gives the optimal measurement operator for the
successful outcome
\begin{equation*}
  M_1 = 
  \begin{pmatrix}
    \tan\theta & 0 \\ 0 & 1
  \end{pmatrix}
  = \tan\theta\ket{0}\bra{0} + \ket{1}\bra{1}
  .
\end{equation*}
The remaining set of matrices representing the measurement operators
need to have a vanishing element to the lower right, and the remaining
elements need to be such that the completeness relation is
satisfied. The easiest way to do this is to simply add one more
operator, namely
\begin{equation*}
  M_2 = 
  \begin{pmatrix}
    \sqrt{1-\tan^2\theta} & 0 \\ 0 & 0
  \end{pmatrix}
  = \sqrt{1-\tan^2\theta} \ket{0}\bra{0}
  .
\end{equation*}
The new states after the measurement are proportional to
\begin{align*}
  M_1\ket{\psi_1} &= \sin\theta\ket{00} + \sin\theta\ket{11}\\
  M_2\ket{\psi_1} &= \cos\theta\sqrt{1-\tan^2\theta}\ket{00}
  .
\end{align*}
The probability for the two outcomes are
\begin{align*}
  p(1)
  &= \bra{\psi_1}M_1^\dagger M_1\ket{\psi_1}
  = \sin^2(\braket{00|00} + \braket{11|11})
  = 2\sin^2 \theta \\
  p(2)
  &= \bra{\psi_1}M_2^\dagger M_2\ket{\psi_1}
  = \cos^2\theta(1-\tan^2\theta)
  = \cos^2\theta - \sin^2\theta
  = 1 - 2\sin^2\theta
\end{align*}
and the normalised new states are
\begin{align*}
  \ket{\psi_{\text{out},1}} &= \frac{1}{\sqrt{2}} (\ket{00} + \ket{11})
  = \ket{\Phi^+}\\
  \ket{\psi_{\text{out},2}} &= \ket{00}
  .
\end{align*}
We see that there is a certain probability for converting a partially
entangled state into a maximally entangled state, but the less
entangled the initial state is, the smaller is the probability. This
is consistent with the fact that LOCC cannot increase the expected
entanglement. Still, we may perform a measurement which may give us a
more entangled state at the risk of losing the entanglement we
had. If we have a large number of bipartite two-qubit systems, we
may apply the above procedure and discard the systems where the
output is a product state, thereby effectively \emph{concentrating}
entanglement. 

% Distilling (and diluting) mixed states
\subsection{Distilling entanglement from mixed states}
\label{sec:distilling_mixed}
\index{entanglement!distillation|(}
Also mixed state entanglement may be concentrated, or
\newconcept{distilled}. This will be important
for practical applications, as the channels used to create
entanglement in general will be noisy.
A collection of distillation protocols was introduced in
\cite{bennett1:1996} and \cite{bennett3:1996}. We will take a closer
look at a slightly modified version of a distillation protocol
introduced in the latter, which for
certain mixed states can yield pure Bell states as output.

Consider a mixed state, as always shared between Alice and Bob,
which can be represented as a collection
of $\ket{\Phi^+}$ states contaminated with a portion of
$\ket{01}$ states. The corresponding density operator is then
\begin{align*}
  \rho &= (1-p)\ket{01}\bra{01} + p \ket{\Phi^+}\bra{\Phi^+} \\
  &= (1-p)\ket{01}\bra{01} + \frac{p}{2}\ket{00}\bra{00}
  + \frac{p}{2}\ket{00}\bra{11} + \frac{p}{2}\ket{11}\bra{00}
  + \frac{p}{2}\ket{11}\bra{11}
  .
\end{align*}
Alice and Bob now pick two pairs from the ensemble. Each of them will
perform a joint operation on the two subsystems on their side. Then
they measure one of the pairs, thereby destroying any entanglement in
that pair, to decide if the other pair is in the desired output state.

The composite system of the two pairs is described by the density
operator
{\footnotesize
\begin{multline*}
  \rho^{\otimes 2}= \\ 
  (1-p)^2\ket{0101}\bra{0101}
    + \tfrac{p(1-p)}{2}\ket{0100}\bra{0100}
    + \tfrac{p(1-p)}{2}\ket{0100}\bra{0111}
    + \tfrac{p(1-p)}{2}\ket{0111}\bra{0100}
    + \tfrac{p(1-p)}{2}\ket{0111}\bra{0111}\\
  + \tfrac{p(1-p)}{2}\ket{0001}\bra{0001}
  \shoveright{
  + \tfrac{p^2}{4}\ket{0000}\bra{0000}
    + \tfrac{p^2}{4}\ket{0000}\bra{0011}
    + \tfrac{p^2}{4}\ket{0011}\bra{0000}
    + \tfrac{p^2}{4}\ket{0011}\bra{0011}}\\
  + \tfrac{p(1-p)}{2}\ket{0001}\bra{1101}
  \shoveright{
    + \tfrac{p^2}{4}\ket{0000}\bra{1100}
    + \tfrac{p^2}{4}\ket{0000}\bra{1111}
    + \tfrac{p^2}{4}\ket{0011}\bra{1100}
    + \tfrac{p^2}{4}\ket{0011}\bra{1111}}\\
  + \tfrac{p(1-p)}{2}\ket{1101}\bra{0001}
  \shoveright{
    + \tfrac{p^2}{4}\ket{1100}\bra{0000}
    + \tfrac{p^2}{4}\ket{1100}\bra{0011}
    + \tfrac{p^2}{4}\ket{1111}\bra{0000}
    + \tfrac{p^2}{4}\ket{1111}\bra{0011}}\\
  + \tfrac{p(1-p)}{2}\ket{1101}\bra{1101}
    + \tfrac{p^2}{4}\ket{1100}\bra{1100}
    + \tfrac{p^2}{4}\ket{1100}\bra{1111}
    + \tfrac{p^2}{4}\ket{1111}\bra{1100}
    + \tfrac{p^2}{4}\ket{1111}\bra{1111}
  \end{multline*}}
where the first and third qubit are in Alice's possession and the
second and fourth are in Bob's possession. The two first qubits are
the first pair, and the two last are the second pair. It may be
written symbolically as $\ket{A_1B_1A_2B_2}$.
Both Alice and Bob applies a CNOT gate to the two qubits in their
possession, using the qubit from the first pair as a control bit and
the second one as the target. A CNOT gate is a unitary transformation
which does nothing when the control bit is zero and flips the target
bit if the control bit is set to 1. The operator is written
\begin{align*}
  U_{\text{CNOT}} &= \ket{00}\bra{00} + \ket{01}\bra{01}
  + \ket{10}\bra{11} + \ket{11}\bra{10}\\
  &= \ket{0}\bra{0} \otimes \unity + \ket{1}\bra{1} \otimes X
\end{align*}
where $X$ is the usual bit flip operator, or Pauli $X$-operator.
But as Alice's operator works on the first and third qubit while Bob's
operator works on the second and fourth, in our basis the operators
are
%\begin{align*}
%  U_{\text{CNOT}}(1,3)
%  &= \ket{0}\bra{0} \otimes \unity \otimes \unity \otimes \unity
%  + \ket{1}\bra{1} \otimes \unity \otimes X \otimes \unity\\
%  U_{\text{CNOT}}(2,4)
%  &= \unity \otimes \ket{0}\bra{0} \otimes \unity \otimes \unity
%  + \unity \otimes \ket{1}\bra{1} \otimes \unity \otimes X
%  .
%\end{align*}
%
% A horribly ugly way to make something look readable:
\begin{equation*}
  \begin{matrix}
    U_{\text{CNOT}}(1,3) & = & \ket{0}\bra{0}\\
    U_{\text{CNOT}}(2,4) & = & \unity
  \end{matrix}
  \;
  \begin{matrix}
    \otimes \\ \otimes
  \end{matrix}\;
  \begin{matrix}
    \unity \\
    \ket{0}\bra{0} 
  \end{matrix}\;
  \begin{matrix}
    \otimes \\ \otimes
  \end{matrix}\;
  \begin{matrix}
    \unity \\
    \unity 
  \end{matrix}\;
  \begin{matrix}
    \otimes \\ \otimes
  \end{matrix}\;
  \begin{matrix}
    \unity  & + & \ket{1}\bra{1}\\
    \unity  & + & \unity
  \end{matrix}\;
  \begin{matrix}
    \otimes \\ \otimes
  \end{matrix}\;
  \begin{matrix}
    \unity \\
    \ket{1}\bra{1}
  \end{matrix}\;
  \begin{matrix}
    \otimes \\ \otimes
  \end{matrix}\;
  \begin{matrix}
    X \\ \unity
  \end{matrix}\;
  \begin{matrix}
    \otimes \\ \otimes
  \end{matrix}\;
  \begin{matrix}
    \unity \\ X
  \end{matrix}
\end{equation*}
Operationally it only amounts to flipping the third qubit of
$\rho^{\otimes 2}$ in all terms where the first qubit is 1 and
flipping the fourth qubit where the second qubit is 1. This leaves the
new state
{\footnotesize
\begin{multline*}
  \rho_{\text{out}} =
  U_{\text{CNOT}}(1,3) U_{\text{CNOT}}(2,4)
  \rho^{\otimes 2}
  U_{\text{CNOT}}^\dagger(2,4) U_{\text{CNOT}}^\dagger(1,3)
  = \\ 
  (1-p)^2\ket{0100}\bra{0100}
    + \tfrac{p(1-p)}{2}\ket{0101}\bra{0101}
    + \tfrac{p(1-p)}{2}\ket{0101}\bra{0110}
    + \tfrac{p(1-p)}{2}\ket{0110}\bra{0101}
    + \tfrac{p(1-p)}{2}\ket{0110}\bra{0110}\\
  + \tfrac{p(1-p)}{2}\ket{0001}\bra{0001}
  \shoveright{
    + \tfrac{p^2}{4}\ket{0000}\bra{0000}
    + \tfrac{p^2}{4}\ket{0000}\bra{0011}
    + \tfrac{p^2}{4}\ket{0011}\bra{0000}
    + \boxed{\tfrac{p^2}{4}\ket{0011}\bra{0011}}}\\
  + \tfrac{p(1-p)}{2}\ket{0001}\bra{1110}
  \shoveright{
    + \tfrac{p^2}{4}\ket{0000}\bra{1111}
    + \tfrac{p^2}{4}\ket{0000}\bra{1100}
    + \boxed{\tfrac{p^2}{4}\ket{0011}\bra{1111}}
    + \tfrac{p^2}{4}\ket{0011}\bra{1100}}\\
  + \tfrac{p(1-p)}{2}\ket{1110}\bra{0001}
  \shoveright{
    + \tfrac{p^2}{4}\ket{1111}\bra{0000}
    + \boxed{\tfrac{p^2}{4}\ket{1111}\bra{0011}}
    + \tfrac{p^2}{4}\ket{1100}\bra{0000}
    + \tfrac{p^2}{4}\ket{1100}\bra{0011}}\\
  + \tfrac{p(1-p)}{2}\ket{1110}\bra{1110}
    + \boxed{\tfrac{p^2}{4}\ket{1111}\bra{1111}}
    + \tfrac{p^2}{4}\ket{1111}\bra{1100}
    + \tfrac{p^2}{4}\ket{1100}\bra{1111}
    + \tfrac{p^2}{4}\ket{1100}\bra{1100}
  .
\end{multline*}}
In this state Alice and Bob both measure the state of their qubits belonging to the
target pair. This can be done formally with the set of measurement operators
\begin{subequations}
  \begin{align}
    M_{00} &= \unity \otimes \unity \otimes \ket{0}\bra{0} \otimes \ket{0}\bra{0}\\
    M_{01} &= \unity \otimes \unity \otimes \ket{0}\bra{0} \otimes \ket{1}\bra{1}\\
    M_{10} &= \unity \otimes \unity \otimes \ket{1}\bra{1} \otimes \ket{0}\bra{0}\\
    M_{11} &= \unity \otimes \unity \otimes \ket{1}\bra{1} \otimes \ket{1}\bra{1}
    ,
  \end{align}
\end{subequations}
and is equivalent to measuring the observable
$\unity \otimes \unity \otimes Z_3 \otimes Z_4$ as the measurement operators are
constructed from eigenvectors of this observable. The interesting
outcome is when both qubits are measured to 1.
The probability for that is
\begin{equation}
  \label{eq:distill_probabillity}
  p(11) = \tr\left(M_{11}\rho_{\text{out}}M_{11}^\dagger\right)
\end{equation}
where only the boxed terms of $\rho_{\text{out}}$ survive being
operated on by the $M_{11}$ and of them only the two diagonal ones 
give a nonzero contribution when the trace is taken,
giving a probability of $\frac{p^2}{2}$.
The new state after the measurement is
\begin{align}
  \label{eq:distill_newstate}
  \frac{M_{11}\rho_{\text{out}}M_{11}^\dagger}{p(11)}
  &=\frac{1}{2}\Big(\ket{00}\bra{00} + \ket{00}\bra{11} + \ket{11}\bra{00}
  + \ket{11}\bra{11}\Big)
  \otimes
  \ket{11}\bra{11} \notag \\
  &= \ket{\Phi^+}\bra{\Phi^+} \otimes \ket{11}\bra{11}
  .
\end{align}

So by using two entangled pairs in a mixed state Alice and Bob have obtained a
pure state where one of the pairs is in a Bell state. The probability
for this to happen is $\frac{p^2}{2}$, so on average we get a yield of
$\frac{p^2}{4}$ Bell states per input pair.

\begin{figure}[htbp]
  \centering
  \resizebox{0.6\width}{!}{\includegraphics{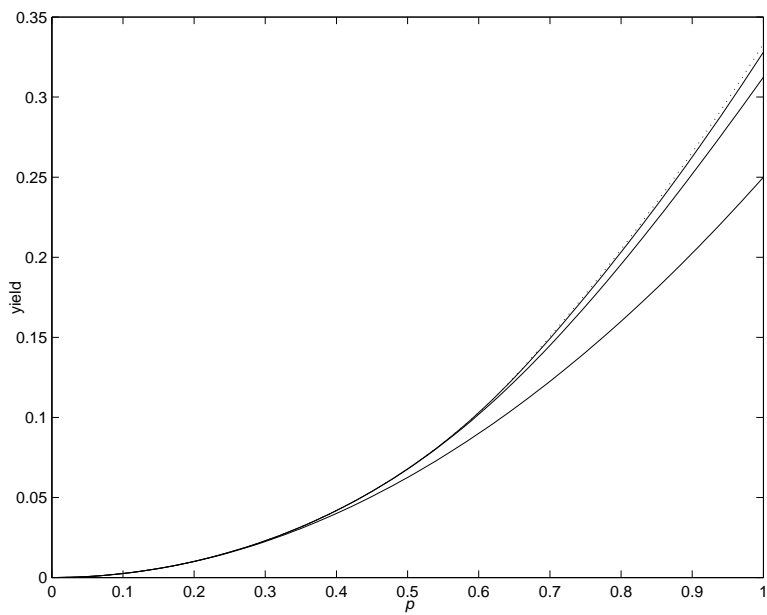}}
  \caption{The yield of the distillation protocol for 1, 2 and 3
  iterations. The dotted line is the asymptotic limit.}
  \label{fig:distill_yield}
\end{figure}

Some additional Bell states may be squeezed out of the mixed states
after using the above protocol. The measurement result 00, with a
probability $\frac{3}{2}p^2 - 2p + 1$, leaves the pair of control
qubits in the state
\begin{equation}
  \label{eq:distill_newstate_00}
  \rho_{00} = \frac{2(1-p)^2}{3p^2 - 4p + 2}\ket{01}\bra{01}
  +
  \frac{p^2}{3p^2 - 4p + 2} \ket{\Phi^+}\bra{\Phi^+}
\end{equation}
which is on the same form as the original input state with a new
mixing ratio
\begin{equation}
  \label{eq:distill_newratio}
  p^\prime = \frac{p^2}{3p^2 - 4p + 2}
  .
\end{equation}
These pairs may be run through the same procedure once more with a
yield of $\frac{{p^\prime}^2}{4}$, increasing the total yield with an
amount of $\frac{1}{2}p(00) \cdot \frac{{p^\prime}^2}{4}$, where the
factor $\frac{1}{2}$ comes from the fact that we use two pairs and
only get one out. This procedure may be iterated, and the total yield
converges quickly so only few iterations are necessary in
practice. Figure \ref{fig:distill_yield} shows the total yield
calculated numerically as a function of $p$. 

In general, distillation protocols do not yield pure state pairs
after a finite number of iterations. Likewise, in the above protocol
this only happens when the state can be written as a mixture of a Bell
state and another special state. A general feature, though, is the
application of the bilateral CNOT unitary transformation on two
pairs.

The purity of the output states is measured by the
\newconcept{fidelity}\index{fidelity} of the state $\rho$,
defined as\footnote{
  Sometimes the fidelity is defined as the square root of this.
}
\begin{equation}
  \label{eq:fidelity}
  F \equiv \bra{\Phi^+} \rho \ket{\Phi^+}
\end{equation}
where $\ket{\Phi^+}$ is the desired output state. Applying a
distillation protocol to an ensemble of states, leaves a subensemble
where the fidelity is higher, along with a subensemble where the
fidelity is lower, often zero. In the limit of reapplying the procedure
to the higher-fidelity states an infinite number of times, the
fidelity approaches unity.
This comes at a cost, however, as the yield decreases as
we demand higher fidelity.
For the simplest protocols the yield approaches zero as the fidelity
approaches unity.
But by combining protocols
and using the already distilled states as a resource, the limit can be
made finite. 
\index{entanglement!distillation|)}

% Maybe some generalisation in the end of this section.
% All operations may be described as completely positive,
% trace preserving superoperators (B Schumacher ``Sending entanglement
% through noisy quantum channels'', Phys. Rev. A 54(4), 2614-2628
% (1996) and K Kraus ``General State Changes in Quantum Theory'',
% Ann. of Phys. 64, 311-335 (1971)

\chapter{Characterisation of bipartite entanglement}
\label{ch:char}

Before moving on to quantify the entanglement in a bipartite state, we
will try to characterise it more qualitatively. There are a number of
questions which does not have an immediate answer.
Given a state, how can we tell wheter it is entangled?
Is there more than one type of entanglement?
Can all entangled states be distilled to Bell states?

\section{Pure states}

When we are given a bipartite state the first question of interest is
whether it is entangled or not. This is the question of
\newconcept{separability}\index{separability}. A
separable state\index{separable state} is a state which is not
entangled, so each of the subsystems can be given
a description on its own.
When it comes to pure states, a state is separable if and only if it
is a product state\index{product state}. As indicated in 
\ref{sec:prelim_entangled}, a pure state is a product state if it can
be written as
\begin{equation}
  \label{eq:product_state}
  \ket{\Psi_\text{prod}} = \ket{\psi}_A \otimes \ket{\phi}_B
\end{equation}
for any pure states $\ket{\psi}_A$ and $\ket{\phi}_B$ belonging to
Alice and Bob, respectively. If the state is not a product state, it
is an entangled state.

In practice, though, one needs to find the right basis vectors to be
able to write a product state as in \eqref{eq:product_state}. A simpler
criterion is desired. The Schmidt decomposition (section
\ref{sec:schmidt}) is helpful in this respect. The number of terms in
the Schmidt decomposition is called the
\newconcept{Schmidt number}\index{Schmidt number}, sometimes also
\newconcept{Schmidt rank}\index{Schmidt rank}.
That is, the number $n$ in
\begin{equation}
  \label{eq:schmidt_number}
  \ket{\psi} = \sum_{k=1}^n
  \lambda_k \ket{a_k^\prime} \otimes \ket{b_k^\prime}
  .
\end{equation}
The Schmidt number is at most $\min(d_A,d_B)$, where $d_A$ ($d_B$) is
the dimension of subsystem $A$ ($B$). But when some of the Schmidt
coefficients are zero, those terms may be skipped giving a lower
Schmidt number. Remember that the Schmidt coefficients are simply the
square root of the eigenvalues of any of the reduced density
matrices. Hence, the Schmidt number is simply the number of nonzero
eigenvalues of $\rho_A$ or $\rho_B$.

What the Schmidt number tells us physically is basically how many degrees of
freedom that are entangled. And as a product state can be written as
\eqref{eq:product_state} it is obvious that its Schmidt number is
1. For pure states this is a necessary and sufficient criterion for
separability.

It turns out that entanglement from any pure entangled state can be
concentrated into Bell states. This may be done by using the procedure
similar to the one described in section \ref{sec:concentrating}, or a
more advanced procedure that operates on more partially entangled
pairs at the same time to give a higher yield (see
e.g.~\cite{bennett2:1996}). Bell states may again be used to produce
other entangled states by means of LOCC. Because of this, and the easy
identification through the Schmidt number, the only
classification we make for bipartite pure states is in separable and
entangled states.

\section{Mixed states}

Some properties of mixed states were discussed in section
\ref{sec:mixed}. The structure of mixed state bipartite entanglement
is richer than its pure state equivalent. For instance there are
entangled states that cannot be distilled to Bell states, and cannot
be put to use in any of the applications where entanglement is a vital
resource. Also, the more difficult task of finding out whether a given
mixed state is separable or not gives rise to classes of states which
can be identified by other criteria.

\subsection{Separable states}
\label{sec:sep_states}

Recall from section \ref{sec:mixed} that a bipartite state is called
separable if it can be realised as a mixture of product states,
\begin{equation}
  \label{eq:def_separable_char}
  \sum_i p_i \, \rho_i^{(A)} \otimes \rho_i^{(B)}
\end{equation}
with $p_i \geq 0$ and $\sum_i p_i = 1$.
Given a density matrix in some basis the task of finding local density
matrices to satisfy \eqref{eq:def_separable_char} can be enormous.
Even for the simplest system which may be entangled, a pair of qubits,
a density operator would typically look like
\begin{equation}
  \begin{aligned}
    \rho &=  
    \tfrac{3}{24}             \ket{00}\bra{00}
    + \tfrac{\sqrt{2}}{24}\im  \ket{00}\bra{01}
    + \tfrac{\sqrt{2}}{12}\im  \ket{00}\bra{11}\\
    &-\tfrac{\sqrt{2}}{24}\im  \ket{01}\bra{00}
    + \tfrac{1}{4}             \ket{01}\bra{01}
    -\tfrac{\sqrt{2}}{12}\im   \ket{01}\bra{10}\\
    &+\tfrac{\sqrt{2}}{12}\im  \ket{10}\bra{01}
    + \tfrac{5}{24}            \ket{10}\bra{10}
    - \tfrac{\sqrt{2}}{24}\im  \ket{10}\bra{11}\\
    &-\tfrac{\sqrt{2}}{12}\im  \ket{11}\bra{00}
    + \tfrac{\sqrt{2}}{24}\im  \ket{11}\bra{10}
    + \tfrac{5}{12}            \ket{11}\bra{11}
  \end{aligned}
\end{equation}
or in matrix representation in the
$\{\ket{00},\ket{01},\ket{10},\ket{11}\}$ basis,
\begin{equation}
  \label{eq:sepexample_matrix}
  \rho =
  \begin{pmatrix}
    \tfrac{3}{24} &
    \tfrac{\sqrt{2}}{24}\im &
    0 &
    \tfrac{\sqrt{2}}{12}\im  \\
    -\tfrac{\sqrt{2}}{24}\im &
    \tfrac{1}{4} &
    -\tfrac{\sqrt{2}}{12}\im &
    0 \\
    0 &
    \tfrac{\sqrt{2}}{12}\im &
    \tfrac{5}{24} &
    - \tfrac{\sqrt{2}}{24}\im \\
    -\tfrac{\sqrt{2}}{12}\im &
    0 &
    \tfrac{\sqrt{2}}{24}\im  &
    \tfrac{5}{12}
  \end{pmatrix}
  .
\end{equation}
It is not easy to see that this can be written as
\eqref{eq:def_separable_char}. Still, it is constructed as a
mixture of two product states.
It can be written as
\begin{align}
  \rho
  &= p_1 \rho_1^{(A)} \otimes \rho_1^{(B)}
  + p_2 \rho_2^{(A)} \otimes \rho_2^{(B)} \notag \\
  &= \frac{1}{2}\left(
    \frac{1}{2} \ket{1}\bra{1} + \frac{1}{2} \ket{-}\bra{-}
  \right)
  \otimes
  \ket{\psi_{\frac{2}{3}}^+} \bra{\psi_{\frac{2}{3}}^+}
  +
  \frac{1}{2}\left(
    \frac{1}{2} \unity/2 + \frac{1}{2} \ket{+}\bra{+}
  \right)
  \otimes
  \ket{\psi_{\frac{2}{3}}^-} \bra{\psi_{\frac{2}{3}}^-}
\end{align}
where
\begin{equation*}
  \ket{\psi_{\frac{2}{3}}^\pm}
  = \frac{1}{\sqrt{3}} \ket{0}
  \pm \sqrt{\frac{2}{3}}\im \ket{1}
\end{equation*}
It could of course be realised by a variety of other mixtures as
well. 

This example shows that we could well use an operational criterion for
separability, like the Schmidt number for pure states.
In general, no easily computable necessary and sufficient criterion is
known. Still, we have a necessary condition which is easy to compute.
This is known as the
\newconcept{PPT\footnote{
    Positive Partial Transpose
  }-criterion}\index{PPT-criterion} or
\newconcept{Peres-Horodecki criterion}
\index{Peres-Horodecki criterion} \cite{PPT:1996,HHH:1996}.
It states that if the state is separable, then the partial transpose
of the density operator with respect to one subsystem is positive. A
transpose of an operator needs to be taken with respect to a basis,
and the resulting operator depends on the basis.
Actually, if $A^T$ is
the transpose of the operator with respect to one basis, any operator
of the form $UA^T U^\dag$ for unitary $U$ satisfying $U = U^T$
(in the same basis)
is a transpose of the operator with respect to
another basis.
However, since the different bases are connected by a unitary
transformation, the eigenvalues are independent of the
basis. Therefore, any orthonormal basis will do for taking the transposition.
More formally, if we choose an orthonormal product basis
$\{\ket{v_i v_j}\} \equiv \{\ket{v_i} \otimes \ket{v_j}\}$
for the state $\rho$, the
partial transpose $\rho^{T_B}$ is defined by its matrix elements
\begin{equation}
  \rho_{m\mu,n\nu}^{T_B}
  = \bra{v_m v_{\mu}}\rho^{T_B}\ket{v_n v_{\nu}}
  = \rho_{m\nu,n\mu}
  .
\end{equation}
In the example matrix \eqref{eq:sepexample_matrix} taking the partial
transpose of system $B$ corresponds to transposing within each of
the four $2 \times 2$ blocks, whereas switching the upper right block with the
lower left block corresponds to transposing system $A$.

For systems of dimension $2 \times 2$ and $2 \times 3$ the PPT-criterion is also
sufficient for the state to be separable \cite{HHH:1996}. For larger
systems, though, there exist entangled PPT states \cite{entangled-ppt}.
The PPT-criterion is still so useful
that the class of PPT states is an important class of states, strictly
larger than the class of separable states. Anything that can be proved
for all PPT states, is automatically true for separable states.

\subsection{Free and bound entanglement}
\label{sec:free_and_bound}
Unlike pure states, not all entangled mixed states can be
distilled into Bell states \cite{bound-entanglement}.
This is an important distinction, because
for many applications entanglement must be distilled in order to be
useful.
Entanglement that can be distilled is called
\newconcept{free entanglement}\index{free entanglement}
and entanglement that cannot be distilled is called
\newconcept{bound entanglement}\index{bound entanglement}. 
Bound entangled states cannot be written as a convex combination of
product states \eqref{eq:def_separable_char} and thus cannot be
prepared locally by two classically communicating parties, but its
entanglement is well hidden and difficult to detect.
It has been shown that for a pair of qubits all entangled states can
be distilled \cite{2x2-distillable}.

A connection between free entangled states and PPT states was found in
\cite{bound-entanglement}. It says that no free entangled state is
PPT. Hence, all PPT entangled states are bound entangled states. This
result was actually the first evidence of the existence of
bound entangled states. 

This raises the question of whether a nonpositive partial transpose
(NPPT\index{NPPT})
state is always distillable. The answer to this question is not
known, but there are indications that NPPT bound entangled states do
exist \cite{bound-NPPT1,bound-NPPT2}.

It has been proven that bound entanglement is useless 
for quantum teleportation \cite{bound_useless_teleportation}, and it
can be proved for other applications as well. One has found, though,
that bound entanglement is not completely useless after all. For
certain tasks it can be activated \cite{bound_activated} to facilitate
distillation of other quantum states. It has even been shown that a
small amount of PPT bound entanglement as a resource makes it possible
to distill entanglement from any NPPT
state \cite{bound_activating_infinitesimal}. 
This means that if NPPT bound entangled states really exist,
entanglement can be distilled by putting together  two states that
are not distillable by themselves.

\chapter{Measures of bipartite entanglement}
\label{ch:measures}

Entanglement can be utilised for performing many tasks which are
impossible without it, or enhance the performance of other tasks. In
most of these tasks entanglement is consumed, so we trade entanglement
for something else. It is clear from this that we may treat
entanglement as a \newconcept{resource}, just as energy is a resource
needed for certain tasks. This is the reason why we want to
\emph{quantify} the entanglement in an entangled state. Given a state
and a task that consumes entanglement, how much can we achieve? How
well can we do? It is not obvious that an entangled state that performs a given
task better than another is will be the better resource for another
task. Indeed, this is only true in a very limited context. Because of
this, we have in general many ways to quantify the nonlocal resources
(or entanglement) in a quantum state.

An entanglement measure\index{entanglement!measure}
$E(\cdot)$ is a functional that takes a quantum
state of a multipartite system to a nonnegative real number. In this chapter we
only consider bipartite systems, hence
\begin{equation}
  \label{eq:def_measure}
  E: \mathcal{D}(\mathcal{H}) \rightarrow \mathbb{R}^+ 
\end{equation}
where $\mathcal{D}(\mathcal{H})$ is the set of density operators on the
Hilbert space $\mathcal{H} = \mathcal{H}_A \otimes \mathcal{H}_B$.
There are in principle two ways to quantify the entanglement in a
state. \newconcept{Operational measures}\index{operational measure}
are based on how well a certain task can be performed, usually
compared to Bell states. The other way, which gives rise to
\newconcept{abstract measures}\index{abstract measure}, is to work
from a set of natural axioms we believe an entanglement measure should
satisfy, and look for functionals that satisfy the axioms. 

% Finite vs. asymptotic regime
There are also two regimes to consider. In the
\newconcept{finite regime}\index{finite regime} we consider the
resources contained in a single quantum system, whereas in the
\newconcept{asymptotic regime}\index{asymptotic regime} we take into
account an infinite number of systems in the same state and
consider the the resources \emph{per system}. 

\section{Pure states}

Before treating the general case of mixed states we will consider the
simpler case of entanglement measures of pure states. The treatment is
facilitated by the fact that a pure state contains no classical
correlations between the subsystems, so any correlation present must
be of quantum nature. As mentioned in section \ref{sec:mixed}, even if
an entangled state is pure the states of the subsystems --
described locally by the
reduced density operator -- are
mixed. For pure states, then, the amount of ``mixedness'' turn out to
be a good measure of entanglement.

\subsection{von Neumann entropy}
\index{von Neumann entropy}
Our measure for how mixed a quantum state $\rho$ is, will be the
\newconcept{von Neumann entropy}, defined as
\begin{equation}
  \label{eq:von_neumann_entropy}
  S(\rho) \equiv - \tr(\rho \log \rho)
\end{equation}
where we take the logarithm base-2 as is the custom in information theory.
It is most easily calculated from the nonzero eigenvalues
$\lambda_i$ of $\rho$ as
\begin{equation}
  \label{eq:von_neumann_entropy_eigen}
  S(\rho) = - \sum_i \lambda_i \log \lambda_i
  .
\end{equation}

The von Neumann entropy is often seen as a generalisation of the
Shannon entropy\index{Shannon entropy}
from classical information theory.
Although it is true
that Shannon entropy arises as a special case in quantum information
theory when only orthogonal states are considered, the historical influence
is the other way around. von Neumann introduced his entropy in 1932
\cite{von_neumann:1932} and Shannon published his mathematical theory
of communication in 1948 \cite{shannon:1948}. In fact, it was von
Neumann who suggested that Shannon should name his uncertainty
function ``entropy'' as that was already used in statistical
physics \cite[p.~180]{tribus:1971}.

What the von Neumann entropy essentially describes,
is the uncertainty in a quantum state. It is zero for pure states, and
smaller for a mixture of two nonorthogonal states than a similar
mixture of two orthogonal states. It is not only an analogue to the
entropy known from thermodynamics, it is the same quantity (up to a
constant with the definition \eqref{eq:von_neumann_entropy}). Indeed,
it was derived using notions such as the ideal gas law and other 
laws of thermodynamics \cite[p. 191-202]{von_neumann:1932}\footnote{
  Pages 359-379 in the far worse typeset English translation
  \cite{von_neumann_english:1955}.
}.

\subsection{Reduced von Neumann entropy -- entropy of entanglement}

For the purpose of quantifying entanglement, the important quantity
is the
\newconcept{reduced} von Neumann entropy.
\index{von Neumann entropy!reduced}
This is the von Neumann entropy of the reduced
density matrix. We saw an example in section \eqref{sec:mixed} that
the reduced density operator of a pure entangled state represents a
mixed state. The von Neumann entropy of \emph{either} of the
subsystems (the entropies are equal) is a good measure of entanglement. Because of this, the
reduced von Neumann entropy is also known under the name
\newconcept{entropy of entanglement}\index{entropy of entanglement}
$E_E$.

It is easy to see that
the reduced von Neumann entropy is equal for both reduced density
matrices. The von Neumann entropy only depends on the eigenvalues of
the density matrix \eqref{eq:von_neumann_entropy_eigen} and because of
the existence of the Schmidt decomposition, the eigenvalues are equal,
as was mentioned in section \ref{sec:schmidt}.

We haven't yet pointed
out what constitutes a good measure, but the entropy of entanglement
has some properties we find natural.
(i) It is zero for
any product state, (ii) it is maximal when the reduced density matrix
is completely mixed, e.g.~when the subsystems have no individual
properties (for the degrees of freedom considered) and (iii) it is
invariant under local unitary transformations. The above requirements
were introduced in \cite{schlienz:1995}, but requirements have later
been refined and placed on firmer ground for both pure and mixed
states. 

The entropy of entanglement was first introduced as a measure of
entanglement in \cite{bennett1:1996}. It is an abstract measure in the
sense that it satisfies some requirements, and it does not have an
immediate operational interpretation. We will now take a look at 
two pure state entanglement measures which have a direct
operational meaning.

\subsection{Entanglement cost and distillable entanglement}

We saw in section \ref{sec:concentrating} an example of a
concentration of pure state entanglement. We turned a partly entangled
two-qubit state into a Bell state with a yield depending on how
entangled the initial state was. There are other protocols which give
a higher yield if we have more than one pair to operate on (see
e.g.~\cite{bennett2:1996}). The
\newconcept{distillable entanglement}\index{entanglement!distillable!pure states}
$E_d$
is defined as the maximum yield of Bell states that can be obtained,
optimised over all possible LOCC protocols. 
The distillable entanglement is also sometimes called
\newconcept{entanglement of distillation}

The \newconcept{entanglement cost}\index{entanglement!cost!pure states} 
$E_c$ is the dual to the distillable entanglement. 
Two separated parties cannot prepare an entangled state if they can
only communicate classically. But if they have some entangled state in
some standard form like Bell states, they can convert those into 
the desired entangled state. Thus, the entanglement cost is defined as
the minimum number of Bell states needed to create a given state by
means of LOCC.

The above definitions were implicitly given for the finite regime. It
turns out that in this regime $E_d$ and $E_c$ are hard to calculate
even though the states are pure. What is clear intuitively is that for
any state
\begin{equation}
  \label{eq:ed_leq_ec}
  E_d \leq E_c
  .
\end{equation}
If not, one could create entanglement by means of LOCC
by converting Bell states to a state not satisfying
\eqref{eq:ed_leq_ec} and converting them back again.

In the asymptotic limit, the results are known.
We define the
regularised (or asymptotic) versions of distillable entanglement
and entanglement cost as
\begin{subequations}
  \begin{align}
    \label{eq:regularised}
    E_D(\ket{\psi})
    &\equiv E_d^\infty(\ket{\psi})
    \equiv \lim_{n \rightarrow \infty}
    \frac{E_d(\ket{\psi}^{\otimes n})}{n}
    ;\\
    E_C(\ket{\psi})
    &\equiv E_c^\infty(\ket{\psi})
    \equiv \lim_{n \rightarrow \infty}
    \frac{E_c(\ket{\psi}^{\otimes n})}{n}
    .
  \end{align}
\end{subequations}
In this case, it was shown in \cite{bennett2:1996} that
both the distillable entanglement and entanglement cost is equal to
the entropy of entanglement,
$E_D(\ket{\psi}) = E_C(\ket{\psi}) = E_E(\ket{\psi})$. In fact, if two
states $\ket{\psi_1}$ and $\ket{\psi_2}$ have the same entropy of
entanglement, they can be interconverted with efficiency
approaching unity as $n\rightarrow \infty$. If they do not have the
same entropy of entanglement, $\ket{\psi_1}$ can be converted into
$\ket{\psi_2}$ with asymptotic yield
$E_E(\ket{\psi_1})/E_E(\ket{\psi_2})$.
% This is not very difficult to show. Do it if there is time.
It has even been shown that this conversion can be done such that the
amount of classical communication required per Bell state produced
approaches zero \cite{interconvertible:1999} (it scales as $\sqrt{n}$).

Despite the fact that the entropy of entanglement did not have an
operational interpretation {\it ab initio}, we see that it indeed plays a
fundamental role. 
The fact that the entropy of entanglement can be
asymptotically conserved in the
conversion of states using LOCC, places huge constraints on
other candidates for pure state asymptotic entanglement measures.
A measure would
not live up to our expectations if we could increase it using only
LOCC. If another measure were to give two different values for states
with the same entropy of entanglement, it would be possible increase
it by LOCC by converting the state with lower value to the one
with higher value. The conversion between states with conversion rate
$E_E(\ket{\psi_1})/E_E(\ket{\psi_2})$ imposes even stronger
constraints, so we would expect any asymptotic pure state measure to
coincide with the entropy of entanglement. The exact conditions for
this to happen are discussed in the next section. 

\subsection{The uniqueness theorem for measures of entanglement}

The fact that the pure state asymptotic entanglement cost and distillable
entanglement coincide with the entropy of entanglement is
a special case of a more general property of entanglement measures.
The
\newconcept{uniqueness theorem for entanglement measures}
\index{entanglement!measure!uniqueness theorem}
states that any pure state
measure of entanglement that satisfies certain natural criteria
coincides with the entropy of entanglement. The criteria are considered
to be too strict for the finite regime, but all natural asymptotic
measures satisfy the conditions.

The uniqueness theorem has been developed gradually in the
literature, in the beginning only as an intuitive idea of the same
character as given in the previous section \cite{uniqueness1}. Later,
uniqueness was proved from a set of conditions that measures
should satisfy \cite{uniqueness2,limits:2000}. In \cite{uniqueness3}
the minimal conditions for the theorem to hold were found.
The conditions come in different versions, some weaker than
others.
It turns out that some of the stronger conditions are not necessary for the
theorem, but are in fact implied by the weaker conditions.
The strongest versions are listed first. 

(P0)
If $\ket{\psi}$ is separable (i.e.~a product state), then $E(\ket{\psi})=0$

(P1a) (Normalisation)
For the maximally entangled state in $d \times d$ dimensions
\begin{equation}
  \ket{\Phi_d^+} = \sum_{i=1}^d \frac{1}{\sqrt{d}} \ket{i}_A \otimes \ket{i}_B
\end{equation}
where the two sets $\{\ket{i}_A\}$ and $\{\ket{i}_B\}$ are 
orthonormal bases for the two subsystems, 
$E(\ket{\Phi_d^+}) = \log d$, where we as usual take the logarithm
base-2.

(P1b)
For the Bell state
$\ket{\Phi^+} = \frac{1}{\sqrt{2}}(\ket{00} + \ket{11})$,
$E(\ket{\Phi^+}) = 1$. The previous condition is simply the
generalisation of this.

(P2)
For any operation $\Lambda$ that can be implemented by means of LOCC
and any $\ket{\psi}$ such that $\Lambda(\ket{\psi})$ is a pure state,
$E(\Lambda(\ket{\psi})) \leq E(\ket{\psi})$.

% This is an oddball. It is not critical for the theorem, nor
% necessarily weaker than (P2)
%(P2b)
%$E(\ket{\psi})$ only depends on the nonzero Schmidt coefficients of
%$\ket{\psi}$.

(P3) (Continuity)
Let $\{\ket{\psi_n}\}$ and $\{\ket{\phi_n}\}$ be sequences of pure
bipartite states living on the sequence of Hilbert spaces
$\{\mathcal{H}_n\}$. For all such sequences such that
$\| \ket{\psi_n}\bra{\psi_n} - \ket{\phi_n}\bra{\phi_n} \|_1 \to 0$ where
$\|\cdot\|_1$ is the trace norm
\index{trace norm}
$\|A\|_1 = \tr(\sqrt{A^\dag A})$,
\begin{equation}
  \label{eq:continuity_pure_states}
  \frac{E(\ket{\psi})-E(\ket{\phi})}{1+\log(\dim \mathcal{H}_n)}
  \rightarrow 0
  .
\end{equation}

(P4a) (Weak additivity)
For all pure states $\ket{\psi}$ and $n \geq 1$,
\begin{equation}
  \label{eq:weak_additivity_pure}
  \frac{E(\ket{\psi}^{\otimes n})}{n} = E(\ket{\psi})
\end{equation}

(P4b) (Asymptotic weak additivity)
Given $\epsilon > 0$ and a pure state $\ket{\psi}$, there exists an
integer $N > 0$ such that for all $n \geq N$,
\begin{equation}
  \label{eq:weaker_additivity_pure}
   \frac{E(\ket{\psi}^{\otimes n})}{n} - \epsilon
   \leq
   E(\ket{\psi})
   \leq
   \frac{E(\ket{\psi}^{\otimes n})}{n} + \epsilon
   .
\end{equation}

(P4c) (Existence of a regularisation)
For all bipartite pure states $\ket{\psi}$, the limit
\begin{equation}
  \label{eq:regularisation_pure}
  E^\infty(\ket{\psi}) \equiv \lim_{n\rightarrow \infty}
  \frac{E(\ket{\psi}^{\otimes n})}{n}
\end{equation}
exists.

With these conditions given, the uniqueness theorem for entanglement
measures states \cite{uniqueness3} that for a functional $E$ on pure states, the
following are equivalent:\\
(i) $E$ satisfies (P1b), (P2), (P3) and (P4b).\\
(ii) $E$ satisfies (P0), (P1a), (P2), (P3) and (P4a).\\
(iii) $E$ coincides with the entropy of entanglement $E = E_E$.\\
On the other hand, if $E$ satisfies (P0), (P1a), (P2) and (P3), it
automatically satisfies (P4c), and $E^\infty = E_E$.

(i) represents the weakest set of conditions. When they are
satisfied, the stronger conditions in (ii) are also satisfied. The
connection is that (P4a) and (P4b) turn out to be equivalent ((P4a)
$\Rightarrow$ (P4b) is obvious) and (P1b), (P2) and (P4a) together
implies (P0) and (P1a). The uniqueness for the asymptotic regime,
previously mentioned, is concretised by the last sentence.

\section{Mixed states}

The measures of pure state bipartite entanglement constitute a very limited
subset of the measures of generally mixed states. For mixed states
there is no unique way to quantify the nonlocal quantum resources even
in the asymptotic regime. In the pure state asymptotic limit, the
entropy of entanglement imposed a total order on the set of states. Any
state could be converted into another state with equal or less
entanglement. In the mixed state case there are states where neither can
be converted into the other, giving only a partial order on the set of
states. (For a longer discussion on entanglement measures as ordering
of the states, see \cite{ordering_states}.) This makes the structure
of mixed state entanglement richer than that of pure states.

\subsection{Entanglement cost and distillable entanglement for
  mixed states}
\index{entanglement!cost!mixed states}
\index{entanglement!distillable!mixed states}

The entanglement cost and distillable entanglement can easily be
defined for mixed states in the same way as for pure states. The
entanglement cost $E_c$ is the minimum number of Bell states needed to
produce the state by means of LOCC and the distillable entanglement
$E_d$ is
the maximum number of Bell states that can be distilled by an optimal
LOCC distillation protocol.
However, a perfect conversion between mixed states in the finite
regime is usually not possible, so in the following we will only consider
the asymptotic versions $E_D$ and $E_C$. There is a
tweak, though, compared to the pure state case. As mentioned in
the discussion of entanglement distillation (section
\ref{sec:distilling_mixed}) we do not require the conversion to be
perfect. The only requirement is that the fidelity approaches unity as
$n \to \infty$. 

Note that to produce mixed states from Bell states using perfect LOCC
we will need to discard some information. This can be done by doing a
random local unitary transformation and not recording which one it was, or
by doing a measurement and forgetting the outcome. A third option is to
entangle the local system with an ancilla system and discard the
ancilla. It is not surprising then that the distillable entanglement 
usually is strictly smaller than the entanglement cost. 
Actually, for any measure $E$ which satisfies certain natural
conditions for asymptotic measures, the entanglement cost and
distillable entanglement provide upper and lower bounds
\cite{limits:2000}\footnote{
  In \cite{limits:2000} the upper bound was $E_F$, the regularised
  \newconcept{entanglement of formation}. This has been shown to be
  equal to the asymptotic entanglement cost \cite{cost_eq_formation}.
},
\begin{equation}
  \label{eq:limits_mention}
  E_D \leq E \leq E_C
\end{equation}

The bound entangled states, discussed in section
\ref{sec:free_and_bound}, is an example where the difference between
$E_C$ and $E_D$ is highly visible. $E_D$ is obviously zero, but $E_C$
is finite. This illustrates that one single measure is not enough to
quantify the entanglement resources in a mixed quantum state, even in
the asymptotic limit. 

\subsection{Axioms for abstract measures}
\label{sec:axioms}

The axiomatic approach to quantifying entanglement has been quite
successful in finding measures that quantify certain aspects of
entanglement. During the course of finding the right conditions, some
conditions 
have been discarded as unnecessary and others have been restricted to
a specific regime. The common denominator, though, has always been
that entanglement, irrespective of how we quantify it, should not
increase on average by any LOCC operation, i.e.~it is monotonic
under LOCC. The first attempt was done
by Vedral \etal in \cite{conditions1} and was slightly improved by
some of the same authors in \cite{conditions2}. Vidal
\cite{uniqueness2} argued that the only absolute requirement was
monotonicity under LOCC. This automatically implies other properties
such as nonchange under local unitary transformations, convexity and
constancy on separable states.

The conditions on mixed states discussed here are taken from
\cite{uniqueness3}. The previously discussed conditions on pure states
are specialisations of those.

(E0a)
$E(\rho) = 0$ if and only if $\rho$ is separable. This is a
useful property for measures that satisfy it, but it is too strict
in general. Bound entangled states, for instance, have zero distillable
entanglement, but are not separable. 

(E0b) 
$E(\rho) = 0$ if $\rho$ is separable.

The normalisation conditions are the same as for pure states.

(E1a) (Normalisation)
For the maximally entangled state in $d \times d$ dimensions,
$\ket{\Phi_d^+}\bra{\Phi_d^+}$, where
\begin{equation}
  \ket{\Phi_d^+} = \sum_{i=1}^d \frac{1}{\sqrt{d}} \ket{i}_A \otimes \ket{i}_B
\end{equation}
and the two sets $\{\ket{i}_A\}$ and $\{\ket{i}_B\}$ are
orthonormal bases,
$E(\ket{\Phi_d^+}\bra{\Phi_d^+}) = \log d$.

(E1b)
For a Bell state $\ket{\Phi^+} = \frac{1}{\sqrt{2}}(\ket{00} + \ket{11})$,
$E(\ket{\Phi^+}\bra{\Phi^+}) = 1$.

(E2a) (LOCC monotonicity)
For any LOCC operation $\Lambda$, $E(\Lambda(\rho)) \leq E(\rho)$.

(E2b)
When $\Lambda$ is a strictly local operation which is either unitary
or adds extra dimensions, then $E(\Lambda(\rho)) = E(\rho)$.

(E2c)
When $\Lambda$ is a strictly local unitary operation, then
$E(\Lambda(\rho)) = E(\rho)$. 

(E3a) (Continuity)
Let $\{\rho_n\}$ and $\{\sigma_n\}$ be sequences of bipartite states
living on the sequence of Hilbert spaces $\{\mathcal{H}_n\}$. For all
such sequences such that $\|\rho_n - \sigma_n\|_1 \to 0$ where
$\|\cdot\|_1$ is the trace norm
\index{trace norm}
$\|A\|_1 = \tr(\sqrt{A^\dag A})$,
\begin{equation}
  \label{eq:continuity_mixed_states}
  \frac{E(\rho_n)-E(\sigma_n)}{1+\log(\dim \mathcal{H}_n)}
  \rightarrow 0
  .
\end{equation}

(E3b)
The previous condition (E3a) is weakened by only requiring it to be
satisfied for approximations to pure sates. That is, (E3a) only needs
to hold when the $\rho_n$ are pure.

Many measures satisfy the weak additivity condition, but it is not
considered necessary for a good measure.

(E4a) (Weak additivity)
For all states $\rho$ and $n \geq 1$,
\begin{equation}
  \label{eq:weak_additivity_mixed}
  \frac{E(\rho^{\otimes n})}{n} = E(\rho)
  .
\end{equation}

(E4b) (Asymptotic weak additivity)
Given $\epsilon > 0$ and a state $\rho$, there exists an
integer $N > 0$ such that for all $n \geq N$,
\begin{equation}
  \label{eq:weaker_additivity_mixed}
  \frac{E(\rho^{\otimes n})}{n} - \epsilon
  \leq
  E(\rho)
  \leq
  \frac{E(\rho^{\otimes n})}{n} + \epsilon
  .
\end{equation}

(E5a) (Subadditivity)
For all states $\rho$ and $\sigma$,
\begin{equation}
  \label{eq:subadditivity}
  E(\rho \otimes \sigma) \leq E(\rho) + E(\sigma)
  .
\end{equation}

(E5b) A special case of the above is when $\rho$ and $\sigma$ are
different numbers of the same state tensored together. For all states
$\rho$, and $m,n \geq 1$,
\begin{equation}
  \label{eq:subadditivity_special}
  E(\rho^{\otimes(m+n)}) \leq E(\rho^{\otimes m}) +  E(\rho^{\otimes n})
  .
\end{equation}

(E5c) (Existence of a regularisation)
For all bipartite states $\rho$, the limit
\begin{equation}
  \label{eq:regularisation}
  E^\infty(\rho) \equiv \lim_{n\rightarrow \infty}
  \frac{E(\rho^{\otimes n})}{n} 
\end{equation}
exists. $E^\infty$ is called the regularisation of $E$.
This is the weakest of the additivity conditions, and is usually
satisfied.

(E6a) (Convexity)
Mixing of states does not increase entanglement. For all bipartite
states $\rho$ and $\sigma$,
\begin{equation}
  \label{eq:convexity}
  E(\lambda \rho + (1-\lambda) \sigma) \leq \lambda E(\rho) + (1-\lambda) E(\sigma)
\end{equation}
for $0 \leq \lambda \leq 1$.

Although (E6a) seems very natural as forgetting which of a set of
prepared states you have should not increase entanglement, there are
indications that the (asymptotic) distillable entanglement is not
convex \cite{distillable_not_convex}.  This is based on the fact that
PPT bound entangled states together with NPPT bound entangled sates (if
they exist) can be distilled. Thus, even a mixture of two states with
zero distillable entanglement can be distilled. A weaker condition
which is shown in \cite{uniqueness3} to hold also for the distillable
entanglement is that the convexity need only hold on decompositions
into pure states.

(E6b) 
For any bipartite state and any pure state realisation
$\rho = \sum_i p_i \ket{\psi_i}\bra{\psi_i}$, $p_i \geq 0$ and $\sum_i
p_i = 1$,
\begin{equation}
  \label{eq:convex_pure}
  E(\rho) \leq \sum_i p_i E(\ket{\psi_i}\bra{\psi_i})
  .
\end{equation}

Like for pure states, asymptotic weak additivity is sufficient for
weak additivity, so (E4a) and (E4b) are actually equivalent. The same goes
for the subadditivity conditions (E5a) and (E5b).

\subsection{$E_D$ and $E_C$ as extreme measures}

In \cite{uniqueness3} the exact conditions for $E_D$ and $E_C$ to be
the lower and upper bound for an entanglement measure were
investigated. Three versions of the theorem were found. With different
conditions on the measures either the nonregularised or the
regularised measure would be bounded by $E_D$ and $E_C$.

(i)
For an entanglement measure $E$ satisfying (E1a), (E2a), (E3a) and (E4b)
(and thereby (E4a)), for all states $\rho$,
\begin{equation}
  \label{eq:limits}
  E_D(\rho) \leq E(\rho) \leq E_C(\rho)
  .
\end{equation}
These conditions are very strong, and until recently no function was
known that satisfied them. The ``squashed entanglement''
\cite{squashed}, however, satisfies all conditions.

(ii)
For an entanglement measure $E$ satisfying (E1a), (E2a), (E3a) and (E5c),
then for all states $\rho$, the \emph{regularised} version $E^\infty$ (which
exists by (E5c) and always satisfies (E4a), but not necessarily (E3a))
is bounded by $E_D$ and $E_C$.
\begin{equation}
  \label{eq:limits_reg}
  E_D(\rho) \leq E^\infty(\rho) \leq E_C(\rho)
\end{equation}
These conditions are easier to satisfy, and they are satisfied by both
the entanglement of formation and the relative entropy of entanglement.

(iii)
Let $E$ be an entanglement measure satisfying (E1a), (E2a), (E3b) and
(E6b). Then if weak additivity (E4a) holds,
\begin{equation}
  \label{eq:limits2}
  E_D(\rho) \leq E(\rho) \leq E_C(\rho)
\end{equation}
and if subadditivity (E5a) holds,
\begin{equation}
  \label{eq:limits_reg2}
  E_D(\rho) \leq E^\infty(\rho) \leq E_C(\rho)
  .
\end{equation}

\subsection{Relative entropy of entanglement and other distance based measures}

Along with the first attempts to give realistic conditions for
entanglement measures \cite{conditions1,conditions2},
a class of measures that satisfied
the conditions was introduced. The conditions considered was that the measure
would be zero if and only if the state is separable (E0a), that local
unitary operations leave it constant (E2c) and that its expectation
value does not increase under LOCC (E2a).

The class of measures is based on some distance function (but not necessarily a
metric) on the set of density matrices. The measure is then the
distance from the state in question to the nearest separable state.
That is, for a distance function
$D(\rho,\sigma)$ an entanglement measure corresponding to the distance
function could be defined as
\begin{equation}
  \label{eq:distance_based}
  E(\rho) \equiv \inf_{\sigma \in \mathcal{S}(\mathcal{H})} D(\rho,\sigma)
\end{equation}
where $\mathcal{S}(\mathcal{H})$ is the set of separable states on
$\mathcal{H}$. The conditions imposed on $E$ is now converted into
conditions on $D$.

The quantum
\newconcept{relative entropy}\index{relative entropy}\footnote{
  For reviews of the role of the relative entropy in quantum
  information theory, see \cite{rel_ent_rev1,rel_ent_rev2}.
}
can be used as a
distance function on the set of density operators. It is defined as
\begin{equation}
  \label{eq:relative_entropy}
  S(\rho \| \sigma) \equiv \tr[\rho (\log \rho - \log \sigma)]
  .
\end{equation}
The relative entropy is not a metric and is not even symmetric. It is
nonnegative, and zero only for identical density operators. The same
unitary operation on both states leaves it invariant.
The quantum
relative entropy can be interpreted as a distinguishability of
quantum states. More precisely, suppose we are given
a large (but finite) number $n$ of quantum systems
that are all in the same state, which is either $\rho$
or $\sigma$. Our task is to
perform measurements to infer which.
The probability for inferring from optimal measurements on
the composite system, that the given state is $\rho$ when it
really is $\sigma$ is \cite{rel_ent_ent}
\begin{equation}
  \label{eq:p_mistake}
  P_n(\sigma \to \rho) = 2^{-n S(\rho \| \sigma)}
\end{equation}
(for large $n$).
This expression is not symmetric in $\rho$ and $\sigma$.

To illustrate this, consider the qubit states
\begin{equation*}
  \rho = \ket{0}\bra{0}
  \qquad
  \text{and}
  \qquad
  \sigma = \frac{1}{2} \ket{0}\bra{0} + \frac{1}{2} \ket{1}\bra{1}
  = \frac{1}{2} \unity_2
\end{equation*}
i.e.~one arbitrary pure state (denoted $\ket{0}$) and the completely
mixed state. 
These states are optimally distinguished by measurement if we measure
each particle in the $\{\ket{0},\ket{1}\}$ basis.
In the case that we are performing measurements on $\sigma$, the
probability for measuring $\ket{0}$ and $\ket{1}$ are both
$\frac{1}{2}$. If all measurements give $\ket{0}$, we would wrongly
conclude that the state was $\rho$. The probability for that to happen
is $2^{-n}$. Conversely, if we perform the measurements on systems in
state $\rho$, we will always get the result $\ket{0}$, so the probability
for confusing it with $\sigma$ is zero.

The quantum relative entropy can be calculated from the eigenvalues
and eigenvectors of the density operators as shown in Appendix
\ref{app:qre}. In the above example, it can be calculated from
\eqref{eq:rel_ent_expr_special} giving $S(\rho \| \sigma) = 1$ and
$S(\sigma \| \rho) = +\infty$.  This is consistent with the
probabilities above and \eqref{eq:p_mistake}. It is a general feature
that $S(\sigma \| \rho) = +\infty$ when $\rho$ is a pure state. This
is because for any pure state there is a complete measurement
for which the outcome is 100\% certain, and using this
measurement would make it impossible to confuse it with another state.

This property alone makes the relative entropy with the arguments
reversed, $D(\rho,\sigma) = S(\sigma \| \rho)$, unusable
in \eqref{eq:distance_based}
as a distance function. It would
give infinity for any pure entangled $\rho$.
$S(\rho \| \sigma)$ on the other hand, behaves well for separable
$\sigma$ and the entanglement measure generated by it,
\begin{equation}
  \label{eq:rel_ent_ent}
  E_r(\rho) \equiv \inf_{\sigma \in \mathcal{S}(\mathcal{H})} S(\rho \| \sigma)
  ,
\end{equation}
is called the
\newconcept{relative entropy of entanglement}
\index{relative entropy!of entanglement}
\index{entropy of entanglement!relative}
\cite{conditions2}.
It could be said that it quantifies the unlikelihood for a separable
state to give measurement outcomes consistent with the entangled
state.

From its definition and the fact that $S(\rho\|\sigma) = 0$ only when
$\rho = \sigma$, it is obvious that the relative entropy of
entanglement satisfies (E0a)
($E_r(\rho)=0$ iff $\rho \in \mathcal{S}(\mathcal{H})$).  
It reduces to the entropy of entanglement on pure states
\cite{conditions2}, and therefore satisfies the normalisation
criterion (E1a) and continuity on pure states (E3b) by the uniqueness
theorem. It is nonincreasing under LOCC operations (E2a)
\cite{conditions1}. 
Continuity (E3a) was shown in \cite{rel_ent_ent_cont}. There
have been found counterexamples to weak additivity (E4a)
\cite{measures_symmetry} and thereby also to asymptotic weak
additivity (E4b). Actually, for very high dimension of a certain class of
states, $E_r(\rho) \approx E_r(\rho \otimes \rho)$. 
It can also be shown that it is both subadditive (E5a) and convex
(E6a) \cite{uniqueness3}.

The regularisation (E5c)
\begin{equation}
  \label{eq:rel_ent_ent_reg}
  E_R(\rho) \equiv E_r^\infty
  = \lim_{n \to \infty} \frac{E_r(\rho^{\otimes n})}{n}
\end{equation}
of the relative entropy of entanglement satisfies automatically (E0a),
(E1a), (E2a), (E5a) and (E6a) as those conditions extends to any
regularisation of a measure that satisfies them
\cite{uniqueness3}. In addition, as any
regularisation does, it is satisfies weak additivity (E4a).

% Variations of E_R by considering other classes
Any distance based measure, including the relative entropy of
entanglement, can be varied by considering other classes of states
than the separable ones. What is important is that the class is closed
under LOCC. 
For instance, the set can be extended to PPT
states. This will give a measure which is easier to calculate and
inferior or equal to the version for separable states. For $2\times2$
and $2\times3$ dimensional systems the measures coincide as the PPT is
then equivalent to separability.

% Other distance measures
Even though the quantum relative entropy is by far the most used of
the distance based measures, other distance functions provide usable
entanglement measures as well. Another measure of distinguishability
of quantum states is the
\newconcept{Bures metric}\index{Bures metric}
\cite{bures:1969,hubner:1992,fuchs:1995}, in this context defined as
\cite{conditions2}
\begin{equation}
  \label{eq:bures_metric}
  D_B(\rho \| \sigma) \equiv 2 - 2\sqrt{F(\rho,\sigma)}
\end{equation}
where
\begin{equation}
  \label{eq:trans_prob}
  F(\rho,\sigma) \equiv \left[ \tr \left\{ \left(
        \sqrt{\sigma}\rho \sqrt{\sigma}
      \right)^{1/2} \right\} \right]^2
\end{equation}
is called Uhlmann's transition probability
\index{Uhlmann's transition probability}\cite{uhlmann:1976}.
The Bures metric is a true metric and thus symmetric.
It is a generalisation of the pure state Fubini-Study metric
to mixed states \cite{hubner:1992}.
The
entanglement measure generated by it satisfies (E0a), (E1b) and (E2a).
Unlike the relative entropy of entanglement, however, it
doesn't reduce to the entropy of entanglement on pure
states. Actually, it is smaller than the entropy of entanglement
\cite{conditions2}.

There may be other distance functions that generate entanglement
measures with nice properties, but the two mentioned are the ones used
in practice. The Hilbert-Schmidt distance
\begin{equation}
  \label{eq:d_hs}
  D_{HS}(\rho,\sigma) \equiv \| \rho - \sigma \|_{HS}^2
  = \tr \left[ (\rho - \sigma)^2 \right]
\end{equation}
was suggested as a candidate
\cite{conditions2,witte:1999}. However, it was shown by Ozawa
\cite{ozawa:2000} that the distance violated one of the
criteria it was conjectured to have (nonexpansion under physical
operations), so it did not satisfy the sufficient conditions
for it to be nonincreasing under LOCC. On the other hand, the very
similar trace norm distance
\begin{equation}
  \label{eq:d_tr}
  D_T(\rho,\sigma) \equiv \| \rho - \sigma \|_1
  = \tr \left[ \sqrt{(\rho - \sigma)^2} \right]
\end{equation}
has been shown to generate an entanglement measure which is monotonic
under LOCC \cite{opposite_rel_ent_ent}.

\subsection{Entanglement of formation}
\label{sec:eof}

The \newconcept{entanglement of formation}
\index{entanglement!of formation}
was historically the first
entanglement measure to appear
\cite{bennett3:1996}.
In the literature it is occasionally called
\newconcept{entanglement of creation}.
It was meant to be the asymptotic entanglement cost, but this
interpretation rests on the weak additivity which is strongly conjectured,
but still not proved. However, the regularisation of it has been shown
to be equal to the
asymptotic entanglement cost \cite{cost_eq_formation}.

The entanglement of formation is a straightforward generalisation
of the entropy of entanglement to mixed states.
Remember that in the asymptotic limit, the entanglement cost in Bell
states of preparing a pure state is given by the entropy of
entanglement. Thus, it was natural to define the entanglement of
formation for a pure state as the entropy of entanglement.
For a given ensemble of pure states
$\mathcal{E} = \{p_i, \ket{\psi_i} \}$
the entanglement of formation is the average of
the entropy of entanglement for the states in the ensemble
\begin{equation}
  \label{eq:formation_ensemble}
  E_f(\mathcal{E}) \equiv \sum_i p_i E_E(\ket{\psi_i})
  .
\end{equation}
A mixed state can be realised by a multitude of pure state ensembles,
with different entanglement of formation. As any of those ensembles
realises the mixed state, the natural definition for the entanglement
of formation for a mixed state is the entanglement of formation for
the ``most economic'' ensemble. That is, the entanglement of formation
for a mixed state is defined as
\begin{equation}
  \label{eq:formation}
  E_f(\rho) \equiv \inf_{\mathcal{E}} \sum_i p_i E_E(\ket{\psi_i})
\end{equation}
where the infimum is taken over all ensembles
$\varepsilon = \{p_i, \ket{\psi_i} \}$ that realises the state
$\rho$. 

How come entanglement of formation is not necessarily equal to the
entanglement cost?
It is easy to see that the asymptotic entanglement cost cannot be greater than
the entanglement of formation. Imagine that we are to produce a large number
$n_{\text{out}}$ ($n_{\text{out}} \to \infty$) of some bipartite state
$\rho_{\text{out}}$. We find the pure state realisation
$\{p_i,\ket{\psi_i}\}$ of
$\rho_{\text{out}}$ that has the lowest average entropy of
entanglement (equal to the entanglement of formation).
Then we produce this ensemble by producing the pure states
in amounts corresponding to the probability distribution $\{p_i\}$. In
the asymptotic limit this can be done reversibly. When we mix the
produced states (discard the information saying which is which), we
have a collection of systems in the state $\rho_{\text{out}}$. Since the entropy of
entanglement is conserved in the first step, and $E_E = 1$ for
Bell states, we have produced systems in state $\rho_{\text{out}}$ at a
cost equal to the entanglement of formation.

The question is if we can do better than this. Intuitively the above
is the most economical way, as any other realisation of
$\rho_{\text{out}}$ would give a higher or equal cost. But there is
another way of producing a large number of $\rho_{\text{out}}$. We can
produce $n_{\text{out}}/N$ copies of the state
$\rho_{\text{out}}^{\otimes N}$. It is not known whether the optimal
pure state realisation of this state will have an average entropy of
entanglement $N$ times that of the state $\rho_{\text{out}}$. In other
words it is not known if $E_f$ is weakly additive,
$E_f(\rho) = \frac{1}{N} E_f(\rho^{\otimes N})$. If it is, then
$E_f = E_F \equiv E_f^{\infty}$ which is equal to the asymptotic
entanglement cost $E_C$.
Additivity has been shown for some special classes of
states \cite{benatti:2001,measures_symmetry,vidal:2002,shimono:2002},
but whether it holds in general, is one of the big open
questions in quantum information theory.

% Properties
From the definition of $E_f$, it is easy to see that it satisfies
(E0a) and (E1a). (E0a) by the fact that the optimal pure state
decomposition of a separable states is into a mixture of product
states, (E1a) by the fact that it reduces to the entropy of
entanglement on pure states. LOCC monotonicity (E2a) was shown when
it was introduced \cite{bennett3:1996}, and continuity (E3a) was shown in
\cite{continuity_eof}. Subadditivity (E5a) follows from the discussion
of additivity above, one may use the same type of decomposition for a
tensor product, but one \emph{might} be able to do better, which would give a
lower $E_f$. Convexity (E6a) follows more or less directly from the
definition.

The way the entropy of entanglement was extended to mixed states in
the case of the entanglement of formation, can also be used for other
measures on pure states. Measures constructed this way are called
\newconcept{convex roof measures}\index{convex roof measure}
\cite{uhlmann:1998,mhorodecki:2001}. The method extends a measure of
some set (here pure states) to the convex hull\footnote{
  The convex hull of a set is the set of all elements that
  can be written as a convex combination of the original set. For instance,
  the convex hull of two points is the line connecting them.
}
(here mixed states), where it is the largest function that is convex and
compatible with the measure on the original set.
The method was used in \cite{roof-negativity} construct an
entanglement measure that extends the entanglement measure called
\newconcept{negativity} to mixed states in another
way than the original measure.

A concept related to the entanglement of formation is the
\newconcept{concurrence}\index{concurrence}
\cite{eof:1997,eof:1998}. It is defined for a system of two
qubits. For a general state $\rho$ of two qubits, let $\tilde{\rho}$ be
the spin-flipped state
\begin{equation}
  \label{eq:spin-flipped}
  \tilde{\rho} \equiv (Y \otimes Y)\rho^\ast (Y \otimes Y)
\end{equation}
where the $Y$ is the Pauli Y operator and $\rho^\ast$ is the complex
conjugate of $\rho$, both taken in the standard basis
$\{ \ket{00}, \ket{01}, \ket{10}, \ket{11} \}$. Let the Hermitian
matrix $R$ be defined as
\begin{equation}
  \label{eq:r_matrix}
  R \equiv \sqrt{\sqrt{\rho}\tilde{\rho}\sqrt{\rho}}
\end{equation}
with eigenvalues in decreasing order $\{\lambda_i\}$. The concurrence
is then defined as
\begin{equation}
  \label{eq:def_concurrence}
  C(\rho) \equiv \max\{0,\lambda_1 - \lambda_2 - \lambda_3 - \lambda_4\}
  .
\end{equation}

The concurrence is monotonic under LOCC, and can thus be used as an
entanglement measure for two qubits. The great advantage is that it is
easily computable. But more important is that it is directly related
to the entanglement of formation, providing an explicit formula for
the entanglement of formation in the case of two qubits. Let the
function $\mathcal{E}$ be given by
\begin{equation}
  \label{eq:eof_from_concurrence}
  \mathcal{E}(C) \equiv h\left( \frac{1+\sqrt{1-C^2}}{2} \right)
\end{equation}
where $h$ is the binary entropy function
\begin{equation}
  \label{eq:bin_entropy}
  h(x) \equiv - x \log x - (1-x)\log (1-x)
  .
\end{equation}
Then the entanglement of formation is simply given by
\cite{eof:1997,eof:1998}
\begin{equation}
  \label{eq:eof_from_concurrence_final}
  E_f(\rho) = \mathcal{E}(C(\rho))
  .
\end{equation}

There have been various attempts to generalise the concept of
concurrence. Uhlmann \cite{gen_concurrence} considered general conjugations instead
of the special \eqref{eq:spin-flipped}. This was further generalised
to concurrence vectors by Auderaert \etal \cite{concurrence-vector}.
The previously
mentioned convex roof extended negativity coincides with the
concurrence in the case of two qubits \cite{roof-negativity}, and thus
provides another generalisation. Some of those generalisations
-- along with other aspects of concurrence and entanglement of
formation -- are reviewed in \cite{review_formation}.

% Formulas for EoF (eof:199{7,8}), Terhal and Vollbrecht, prl 85,
% 2625, etc. 

\subsection{Negativity}
\label{sec:negativity}

The entanglement measures discussed so far have all had a serious
drawback. Their definition includes some kind of optimisation, which
make their evaluation very difficult. In fact, expressions for the
relative entropy of entanglement and entanglement of formation has
only been calculated for highly symmetric states
(e.g.~\cite{conditions2,eof_isotropic,measures_symmetry,rel_ent_ent_oo-inv})
and low dimensional cases
(e.g.~\cite{eof:1997,eof:1998}).
For general states, evaluating a measure includes heavy numerical
calculations. With the aim of introducing a \emph{computable} measure
of entanglement, two related quantities were defined by Vidal and
Werner \cite{negativity}. They can both be seen as a quantification of
the PPT criterion for separability, and because of that they have the
disadvantage that they fail to distinguish between separable states and
entangled PPT states. Both quantities are based on the trace norm of
the partial transpose (in some basis) of a state, $\| \rho^{T_B}
\|_1$, which can easily be calculated using standard linear algebra
packages. The first quantity is the
\newconcept{negativity}\index{negativity}
\begin{equation}
  \label{eq:negativity_def}
  \mathcal{N}(\rho) \equiv \frac{\| \rho^{T_B} \|_1 - 1}{2}
  .
\end{equation}
This quantity is equal to the absolute value of the sum of negative
eigenvalues of $\rho^{T_B}$. It was first introduced in
\cite[Appendix B]{volume_separable} (without the factor $\frac{1}{2})$
and later shown to not increase under LOCC
\cite{negativity}. The other quantity is the
\newconcept{logarithmic negativity}\index{negativity!logarithmic}
\begin{equation}
  \label{eq:def_negativity_log}
  E_\mathcal{N}(\rho) \equiv \log \| \rho^{T_B} \|_1
  ,
\end{equation}
which is not strictly monotonic under LOCC, but does not increase for
a subclass of LOCC.

For all separable states (and more generally all PPT states), both
$\mathcal{N}$ and $E_\mathcal{N}$ vanish. This is easily seen from the
fact that the partial transpose operation does not change the trace of
the density operator. If $\rho^{T_B}$ is a positive operator
(i.e~$\rho$ satisfies the PPT criterion for separability), then
$\| \rho^{T_B} \|_1 = 1$, since $\| \rho^{T_B} \|_1$  is the
sum of the absolute values of the eigenvalues of $\rho^{T_B}$
and all eigenvalues are positive. For states violating the
PPT criterion, $\| \rho^{T_B} \|_1 > 1$. In this sense, both
quantities quantify how much the state violates the PPT
criterion.

% Properties
A lot of properties of $\mathcal{N}$ and $E_\mathcal{N}$ were deduced in
\cite{negativity}.
As already mentioned, $\mathcal{N}$ is zero for all separable states,
but is also zero for PPT entangled states. Hence, it satisfies (E0b)
but not (E0a). It is $\frac{1}{2}$ for Bell states, so it doesn't
satisfy the normalisation criteria either. It could be made
to satisfy (E1b) by defining it as the double value, but the
normalisation wouldn't extend to maximally entangled states in
dimension $d \times d$ (E1a). It is monotonic under LOCC (E2a),
convex (E6a), and superadditive, satisfying
$\mathcal{N}(\rho \otimes \sigma)
= \mathcal{N}(\rho) + \mathcal{N}(\sigma)
+ 2 \mathcal{N}(\rho) \mathcal{N}(\sigma)$. Thus, neither additivity
of any kind (E4a) nor subadditivity (E5a) is satisfied.

Like $\mathcal{N}$, logarithmic negativity $E_\mathcal{N}$ satisfies (E0b) and
not (E0a). However, unlike $\mathcal{N}$ it also satisfies the
normalisation condition (E1a). But it does not coincide with the entropy of
entanglement for pure states in general. Actually $E_\mathcal{N} \geq
E_E$ on pure states, with equality only for maximally entangled
states.
It is not monotonic under LOCC (E2a), but is an upper
bound for the distillable entanglement. Nor is it convex (E6a). The
most appealing property of $E_\mathcal{N}$ is probably that it
satisfies (strong) additivity,
$E_\mathcal{N}(\rho \otimes \sigma)
= E_\mathcal{N}(\rho) + E_\mathcal{N}(\sigma)$.
From this, weak additivity (E4a) and subadditivity (E5a) comes for
free. 

% Generalisations
The negativity can be generalised to give rise to other entanglement
measures. This is done by replacing $\| \rho^{T_B} \|_1$ by another norm
$\| \rho \|$. The generalisation comes naturally if we define the
trace norm in another, but equivalent way. Any Hermitian matrix can be
written as a difference of two positive operators,
\begin{equation}
  \label{eq:herm_difference}
  A = a_+ \rho^+ - a_- \rho^-
\end{equation}
where $\rho^\pm$ are density matrices and $a_\pm$ are nonnegative
numbers. There exists a decomposition of the form
\eqref{eq:herm_difference} which minimises $a_+ + a_-$. For this
decomposition the trace norm\index{trace norm} is
$ \| A \|_1 = a_+ + a_-$, and $a_-$ is the absolute sum of the
negative eigenvalues of $A$ \cite{negativity}. From this we get another
expression for the negativity, namely
\begin{equation}
  \label{eq:negativity_var}
  \mathcal{N}(A) = \inf\{a_-\ |\ A^{T_B} = a_+ \rho^+ - a_- \rho^- \}
  .
\end{equation}
Minimising $a_-$ is in this case the same as minimising $a_+ + a_-$,
since by taking the trace of the condition, we get
$a_+ = a_- + \tr(A^{T_B})$. If $A^{T_B}$ is a density matrix, A is
PPT, and we can take $a_-$ to be zero, so the negativity vanishes as
it should for PPT states.

The condition $A^{T_B} = a_+ \rho^+ - a_- \rho^-$ restricts the set of
density matrices $\rho^\pm$ over which we can take the
infimum. Another norm and another negativity can be obtained by taking
another set $S$ from which $\rho^\pm$ must be picked. Norms that can
be defined in this way corresponding to a compact set $S$ are called
\newconcept{base norms}\index{base norm}. The norm corresponding to
a general compact set $S$ is then
\begin{equation}
  \label{eq:base_norm}
  \| A \|_S = \inf\{ a_+ + a_-\  |\  A=a_+\rho^+ - a_-\rho^-,\,
  a_\pm \geq 0, \, \rho^\pm \in S
  \}
\end{equation}
and the corresponding negativity
\begin{equation}
  \label{eq:gen_negativity}
  \mathcal{N}_S(A) =  \inf\{ a_- \ | \ A=a_+\rho^+ - a_-\rho^-,\,
  a_\pm \geq 0, \, \rho^\pm \in S
  \}
  .
\end{equation}
We get back the usual trace norm and negativity from this by taking $S$ as the
set of all Hermitian (not necessarily positive) matrices with unit
trace such that the partial transpose is positive. In mathematical
terms, $S = \{ A \ | \ A = A^\dag,\ \tr(A)=1,\ A^{T_B} \geq 0 \}$.

By choosing $S$ as the set of all separable states, another
entanglement measure arises. It was described prior to the
introduction of negativity and was introduced in
\cite{robustness}. It is called
\newconcept{robustness of entanglement}
\index{robustness of entanglement} and was
defined as the minimum amount of mixing with
locally prepared states needed to make the state separable.
Also other measures have been related to negativities
\cite{mhorodecki:2001}, such as the measures introduced by Rudolph
based on the greatest cross norm \cite{cross_norm}. It should be noted
that only the original negativity has the advantage of being
easily computable. The other negativities need to be evaluated by
finding the infimum over a large parameter space.

\subsection{Squashed entanglement}

Of the entanglement measures considered so far, only few are
additive, and when they are some of the other greatly desired
conditions are not fulfilled. Quite recently a new entanglement
measure was proposed, having more nice properties than its
predecessors. It is motivated from the so called intrinsic information
of classical cryptography and is called
\newconcept{squashed entanglement}\index{entanglement!squashed}
\cite{squashed}. It is defined as
\begin{equation}
  \label{eq:def_squashed}
  E_\text{sq}(\rho) \equiv \inf\left\{
    \frac{1}{2} I(A;B|E) \ |\  \rho^{ABE} \ \text{extension of}\  \rho
    \ \text{to}\ \mathcal{H}_E
  \right\}
\end{equation}
The infimum is taken over all extensions to a third subsystem $E$,
so $\rho \equiv \rho^{AB} = \tr_E(\rho^{ABE})$. $I(A;B|E)$ is the quantum mutual
conditional information defined as
\begin{equation}
  \label{eq:def_qcondmutinfo}
  I(A;B|E) \equiv S(\rho^{AE}) + S(\rho^{BE}) - S(\rho^{ABE}) - S(\rho^E)
  .
\end{equation}

% Properties
Most of the properties of $E_\text{sq}$ were shown along with its
introduction as an entanglement measure in \cite{squashed}.
The squashed entanglement vanishes on all separable states (E0b), but
it is not known whether it also vanishes for some entangled states
(E0a). If there exists an extension to $\mathcal{H}_E$ with $\dim
\mathcal{H}_E < \infty$ for which  $I(A;B|E)$ vanishes, it implies
that $\rho^{AB}$ is separable. But the infimum in
\eqref{eq:def_squashed} is taken over all extensions, and thus
$E_\text{sq}(\rho^{AB})$ may be zero even if no finite dimensional
extension makes $I(A;B|E)$ zero. It coincides with the entropy of
entanglement on pure states, and thus satisfies the normalisation
condition (E1a). It is also bounded from below by the distillable
entanglement and from above by the entanglement cost. As a proper
entanglement measure it does not increase under LOCC (E2a) and it is
convex (E6a). When it was introduced, it was shown to be continuous on
most of the state space, and indications that it was continuous on the
whole state space were presented. Continuity was finally proved by
Alicki and Fannes \cite{squashed_cont} (E3a). The measure is strongly additive,
making it satisfy both weak additivity (E4a) and subadditivity
(E5a). So except from (E0a), which is still unknown, squashed
entanglement satisfies the strictest versions of all conditions
presented.

In \cite{measures_symmetry} various additivity properties were
discussed. The additivities we have presented so far are for
\emph{independent} pairs, described by tensor products. But if there
are entanglement or classical correlations between the pairs as well,
they may be taken advantage of. Consider the state
$\rho^{A B A^\prime B^\prime}$ which lives on the Hilbert space
$\mathcal{H}_A \otimes \mathcal{H}_B \otimes \mathcal{H}_{A^\prime} 
\otimes \mathcal{H}_{B^\prime}$, where $\mathcal{H}_A \otimes
\mathcal{H}_{A^\prime}$ is controlled by Alice and $\mathcal{H}_B
\otimes \mathcal{H}_{B^\prime}$ by Bob. Each of the primed and
unprimed subsystems are individually described by tracing out the
other,
\begin{equation}
  \rho^{AB} = \tr_{A^\prime B^\prime}(\rho^{A B A^\prime B^\prime})
  , \qquad
  \rho^{A^\prime B^\prime} = \tr_{AB}(\rho^{A B A^\prime B^\prime})
  ,
\end{equation}
and if they are independent
$\rho^{A B A^\prime B^\prime}
= \rho^{AB} \otimes \rho^{A^\prime B^\prime}$. For a strongly
additive entanglement measure this implies
$E(\rho^{A B A^\prime B^\prime})
= E(\rho^{AB}) + E(\rho^{A^\prime B^\prime})$. But if there are
correlations between the pairs that can be exploited, we would expect
\begin{equation}
  \label{eq:strong_super}
  E(\rho^{A B A^\prime B^\prime})
  \geq E(\rho^{AB}) + E(\rho^{A^\prime B^\prime})
  .
\end{equation}
This property is called
\newconcept{strong superadditivity}\index{superadditivity!strong}
\cite{measures_symmetry} and is satisfied by the squashed entanglement
\cite{squashed}.

Like most other entanglement measures, the definition of squashed
entanglement involves an optimalisation which makes the measure hard
to compute. Due to its recent introduction, not much work has been
published about its calculation on special sets of states, nor about
efficient numerical algorithms. One of the important open questions is
whether it vanishes on any entangled states, especially on the bound
entangled ones.

\subsection{Other entanglement measures}

In addition to the entanglement measures discussed, other measures
have been proposed without gaining the same importance. One of them is
the \newconcept{entanglement of assistance}\index{entanglement!of assistance}
\cite{cohen:1998,assistance:1998}. It is based on the following idea.
Many copies of an entangled tripartite pure state, $\ket{\Psi^{ABC}}$, is
shared between Alice, Bob and Charlie. Now, Alice
and Bob want to use the state to perform some task involving the use
of bipartite entanglement. If they ignore Charlie completely, they share
the state $\rho^{AB} = \tr_C(\ket{\Psi^{ABC}} \bra{\Psi^{ABC}})$. In
other words they share a mixed state, and 
it might even be separable. But Charlie wants to help, and Alice
and Bob may communicate classically with him. Therefore, Charlie can
perform local operations on his part of the system and tell Alice and
Bob about the measurement outcomes. A measurement by Charlie will then
leave the subsystem of Alice and Bob in some pure state, which depending on the
measurement outcome may or may not be entangled. Alice and Bob can then
discard the pairs that are not entangled, and turn the rest reversibly
(in the asymptotic limit) into Bell states. If each measurement by
Charlie leaves them with a state $\ket{\Psi_i^{AB}}$ with probability
$p_i$, the number of Bell states produced per copy of the tripartite
state will be equal to the
average entropy of entanglement of the ensemble
$\mathcal{E} = \{ p_i, \ket{\Psi_i^{AB}} \}$, namely
\begin{equation}
  \label{eq:assistance1}
  E(\mathcal{E}) = \sum_i p_i E_E(\ket{\Psi_i^{AB}})
.
\end{equation}
Now, Charlie wants to do
the measurements in a way that leaves Alice and Bob with as much
entanglement as possible on average.
It turns out that
the optimisation can be
taken over all ensembles $\mathcal{E} = \{ p_i, \ket{\Psi_i^{AB}} \}$ consistent
with $\rho^{AB}$.
Thus, what Charlie can
achieve does not depend on the initial tripartite state
$\ket{\Psi^{ABC}}$, only on $\rho^{AB}$, so all pure tripartite states
for which $\rho^{AB}$ is the same, allow Charlie to help Alice and Bob
to the same result.
The entanglement of assistance is the average entropy of entanglement
Alice and Bob are left with after Charlie has applied his optimal
strategy, namely
\begin{equation}
  \label{eq:assistance2}
  E_a(\rho^{AB}) \equiv \sup_{\mathcal{E}} \sum_i p_i E_E(\ket{\Psi_i^{AB}})
  .
\end{equation}
We can see that this is dual to the entanglement of formation
\eqref{eq:formation}, which is the same with an infimum instead of the
supremum. Even though the entanglement of formation is believed to be
additive, there are states for which the entanglement of assistance is
superadditive,\index{superadditivity}
\begin{equation}
  \label{eq:assistance_superadd}
  E_a(\rho^{AB} \otimes \sigma^{AB})
  \geq E_a(\rho^{AB}) + E_a(\sigma^{AB})
  .
\end{equation}

A recently introduced measure related to the entanglement of assistance is the
\newconcept{localizable entanglement}
\index{entanglement!localizable}
\cite{localizable:2004}.
It is really a multipartite measure for a large number of qubits, but
the goal is, just as for the entanglement of assistance, to optimise
(localise) the entanglement between two subsystems by performing
measurements on the others. If we had allowed global measurements on
the other qubits, we could have regarded it as a single ``Charlie''
system, and it would have reduced to the entanglement of
assistance. Because we can only perform local measurements on each
individual qubit, the localizable entanglement is necessarily smaller
than the entanglement of assistance.

Another entanglement measure recently introduced is the
\newconcept{witnessed entropy of entanglement}
\index{entropy of entanglement!witnessed}
\cite{witnessed_ent_ent}. It is based on the concept of entanglement
witnesses \cite{HHH:1996}.
An entanglement witness for an entangled state $\sigma$ is a Hermitian operator
$W$ for which $\tr(W\sigma) < 0$ and
$\tr(W\rho) \geq 0$ for all separable states $\rho$.
We conventionally normalise entanglement witnesses to
have $\tr(W)=1$. We say that an entanglement
witness $W_\sigma$ is optimal for a state $\sigma$ if
\begin{equation}
  \label{eq:optimal_witness}
  \tr(W_\sigma \sigma) \leq \tr(W \sigma)
\end{equation}
for all entanglement witnesses $W$. The definition of the witnessed
entropy of entanglement is based on the optimal witness,
\begin{equation}
  \label{eq:def_witnessed_ent_ent}
  E_w(\rho) \equiv \log(D/d) \max[0,-\tr(W_\rho \rho)]
\end{equation}
where $D$ is the dimension of the total Hilbert space and $d$ the
smallest of the dimensions of the Hilbert spaces of the
subsystems. Note that the definition applies equally well to bipartite
and multipartite systems. It is not additive, but satisfies the rest
of our conditions for entanglement measures, (E0a), (E1a), (E2a),
(E3a), (E5a) and (E6a). The main advantage of this measure is that
it can be approximately calculated for all mixed states. 
Finding the exact optimal entanglement witness, is a
hard optimisation problem, but using approximation techniques, there
are algorithms to find the $E_w$
to a given precision. 

Other measures can be defined operationally in the following way. Take
your favourite application of entanglement, study how well or bad it
performs for different sets of states, define an entanglement measure
from this and check that it satisfies at least some of the conditions
for entanglement, above all LOCC monotonicity. One of the operational
measures that have had limited impact on the community is an
entanglement measure introduced by Hiroshima \cite{E_dc} based on
the capacity of dense coding \cite{dense_coding}. Another
operational measure by Biham \etal \cite{E_ga} is
derived from Grover's search algorithm
\cite{grover_proceedings,grover_journal}. It is based on the fact that
the algorithm performs worse the more the input state is
entangled.

\chapter{Beyond bipartite entanglement and finite dimension}
\label{ch:multipartite}

The systems we have considered until now have all been bipartite
systems with finite dimensional Hilbert spaces. Some of the concepts
and measures considered are valid also in the multipartite systems and
for entanglement of continuous variables. Others lose their meaning
in this more general setting. In this chapter we consider some of the steps that need
to be taken to generalise the concepts from the previous sections to
multipartite systems and systems with continuous variables.

\section{Multipartite systems}
\index{multipartite system}

% Introduction with examples
While the basic structure of bipartite entanglement is well
understood, multipartite entanglement is still an active field of
research. In bipartite systems the pure state entanglement in the
asymptotic regime is well understood, since the von Neumann entropy of
the reduced density matrices can be conserved under
transformations. In general the Schmidt decomposition cannot be
extended to multipartite systems \cite{peres_multischmidt}, so the
eigenvalues of the reduced density matrices of each system need not be
the same, nor do the von Neumann entropies.

The questions about entanglement literally acquire new dimensions when
more that two subsystems are considered. Between two parties all
relations are between those two parties, and in terms of quantum
communication there is only one single channel.
Once we increase the number of parties to three, we have three pair
relations in the system. Still, those three pair relations do not
describe all correlations in the system (neither classical nor
quantum). Consider for instance the pure tripartite entangled state
called the GHZ\footnote{
  Daniel M. Greenberger, Michael A. Horne and Anton Zeilinger
}
state\index{GHZ state}
\begin{equation}
  \label{eq:tripart_ex}
  \ket{\mathit{GHZ}} \equiv \frac{1}{\sqrt{2}}(\ket{000} + \ket{111} )
.
\end{equation}
If we trace out any one of the systems, we get the density matrix shared
by the two remaining parties
\begin{equation}
  \label{eq:tripart_ex_dens}
  \rho = \frac{1}{2}(\ket{00}\bra{00} + \ket{11}\bra{11})
\end{equation}
which is separable. Thus, the state contains no bipartite entanglement
between the systems, but the tripartite entanglement is maximal.

% Separability
\subsection{Separability}
Some notions can be taken directly over from bipartite systems,
though. The generalisation of separability from \eqref{eq:def_separable_char} 
is straightforward. A state is separable if and only if it can be
written as a convex combination of product states;
\begin{equation}
  \label{eq:def_sep_multi}
  \sum_i p_i \, \rho_i^{(A)} \otimes \rho_i^{(B)}
  \otimes \cdots \otimes \rho_i^{(N)}
\end{equation}
with $p_i \geq 0$ and $\sum_i p_i = 1$.

The classification of states does not end with this. An entangled
multipartite state may or may not still be entangled after one of the
subsystems are ignored (i.e.~traced out). A multipartite state that loses its
entanglement when any of the subsystems is traced out, is called
\newconcept{multiseparable}\index{multiseparable}
\cite{thapliyal_multisep}.

The PPT criterion for separability trivially extends to the
multipartite case. If a multipartite state is separable, then
transposing any number of the subsystems must also give a valid
(i.e.~positive) density operator.

% Generalisation of measures
\subsection{Measures of entanglement}

% Entanglement cost and distillable entanglement
The definition of entanglement cost and the distillable entanglement is
based on a standard bipartite state, namely the Bell state. This is a
bipartite state, so it is not obvious how to extend these
measures to the multipartite case. Nielsen \cite{ounces_and_pound}
argued that this could by overcome by generalising the definition from
Bell states to a definition in terms of qubits transferred (in the
case of creation) or the number of CNOT gates that could be
implemented (in the case of distillation). For instance, to create a
Bell state it is necessary to transfer one qubit. Likewise
the transfer of one qubit can be simulated by teleporting a state,
spending one Bell state. If the entanglement cost is defined in terms of how
many qubits that need to be transferred, it will coincide with the
usual definition for bipartite states.

% Conditions
Most of the conditions we imposed on measures in section
\ref{sec:axioms} generalise to the multipartite case. The
only possible exceptions are
of course the normalisation conditions (E1a) and (E1b) which are based
on Bell states. Those would be unnatural to impose on multipartite
entanglement, but it should be possible for a measure to satisfy them
for states where only two of the subsystems are entangled.

% Distance based
The relative entropy of entanglement and other distance based
measures do not have any reference to the bipartite case in their
definition. Therefore, they are just as valid in the multipartite
case as in the bipartite case.

% Entanglement of formation
The entanglement of formation suffers the same problem as the
entanglement cost, and can be generalised in the same way. For a pure
state we define the entanglement of formation as the number of qubits
we need to transfer in order to prepare it (LOCC comes for free). Then
we extend the definition to the mixed states by the convex roof method
as usual. Another more formal generalisation was suggested by Wang
\cite{eof_multi}.

The meaning of entanglement measures in the multipartite setting is
more vague than in the bipartite. In the bipartite setting it
quantifies in some way how much that can be achieved by the two
parties together. In the multipartite case there are more
possibilities. For instance there may be some entanglement
between all of the parties, as in the state
\begin{equation}
  \label{eq:w_state}
  \ket{W} \equiv \frac{1}{\sqrt{3}}(\ket{100}+\ket{010}+\ket{001})
\end{equation}
which on tracing out one of the subsystems gives the density matrix
\begin{equation}
  \label{eq:w_tracedout}
  \rho = \frac{1}{3}\ket{00}\bra{00} + \frac{2}{3} \ket{\Psi^+}\bra{\Psi^+}
\end{equation}
which is an entangled state with negativity $\mathcal{N}=0.206$ and
entanglement of formation $E_f = 0.550$ (calculated from the
expressions in sections \ref{sec:negativity} and \ref{sec:eof}). 
Or there may be no entanglement between the pairs, as in the GHZ state
mentioned earlier. 
Both states can be prepared by sending two qubits
(after one of the parties has created the state locally). While the
different bipartite measures place different orderings
on the states \cite{different_ordering}, ordering is even more ambiguous for
multipartite states. The types of entanglement are qualitatively more
different, and a state being very much entangled in one sense (as
defined by one measure), can
have very little or no entanglement in another sense.
All this says is that, since a multipartite system is a more
complicated structure, we cannot expect one single parameter to be as good a
description of the entanglement in it as it is in the bipartite case.

\section{Continuous variables}

Most of quantum information theory has been formulated in the
context of finite dimensional Hilbert spaces. However, from an
experimental point of view, entanglement is just as interesting in the
continuous variable regime. This has led to efforts to quantify
entanglement also in the case where the Hilbert spaces have infinite
dimension. Two papers by Eisert \etal
\cite{contvar:2002,contvar-intro:2003} summarise many of the new
problems that arise and their possible solutions.

First of all, when the dimension of the Hilbert space is infinite, we
have states with an infinite amount of entanglement. This should come
as no surprise, as the entropy of entanglement for a maximally
entangled state in $d \times d$ dimensions is $\log d$
(c.f.~criteria (P1a) and (E1a)). What is more worrying is that the set
of pure states with infinite entropy of entanglement is actually dense in the
trace norm on the set of pure states \cite{contvar:2002}. This means
that arbitrarily close to any product state is a state which has
infinite entanglement. The solution to this problem is to consider
only states where the mean energy is bounded from above. This is a
reasonable physical assumption, and on this subset of states the
entropy of entanglement is continuous. Other measures can also be
defined on this subset without exhibiting strange behaviour.

% Separability
The definition of separability in a continuous variable bipartite
system must be slightly altered from the one in section
\ref{sec:sep_states}. The expression is still the same as
\eqref{eq:def_separable_char}, namely
\begin{equation}
  \label{eq:def_separable_inf}
  \sum_i p_i \, \rho_i^{(A)} \otimes \rho_i^{(B)}
\end{equation}
with $p_i$ is a probability distribution. But in infinite dimension
we can only require that the state can be approximated arbitrarily
well by an expression of the form \eqref{eq:def_separable_inf}.

% Sep. criteria
The PPT criterion for separability is also valid in infinite dimension
and for the special class of so-called Gaussian states, it is also
sufficient for separability \cite{ppt_contvar}.

It is a well known fact in the finite dimensional case that there is a
neighbourhood around the completely mixed state $\unity/d$ where all
states are separable
\cite{volume_separable,sep_fourier,myrh_sep}. However, the volume of
the separable ball diminishes as the dimension of the system grows. 
In the infinite
dimensional setting, even when the energy is bounded from above, it
has been shown \cite{contvar:2002}
that arbitrarily close to any state (in the trace norm),
there is an entangled state.

% Gaussian states
The class of so-called Gaussian states has a special role in
entanglement theory of continuous variables. These are states with a
Gaussian Wigner function (a quasi-probability distribution in phase space which
completely describes a quantum state). These states are often
encountered in the laboratory, and they can be completely described by
two parameters. For Gaussian states, and operators that takes Gaussian
states to Gaussian states, many of the notions from finite dimensional
entanglement theory have similar counterparts. See
e.g.~\cite{contvar-intro:2003,ent_prop_gauss,eof_sym_gauss,gaussian_eof,lower_eof_gauss}.

%appendices
\appendix

\chapter{Expressions for the quantum relative entropy}
\label{app:qre}

The quantum relative entropy is defined as
\begin{equation}
  \label{eq:rel_ent_def_app}
  S(\rho \| \sigma) \equiv
  \tr\{ \rho \log \rho - \rho \log \sigma  \}
  .
\end{equation}
We want to calculate it in terms of the eigenvectors and eigenvalues
of $\rho$ and $\sigma$. We denote the eigenvalues of $\rho$ ($\sigma$)
by $r_i$ ($s_i$) and its eigenvectors by $\ket{r_i}$ ($\ket{s_i}$).
In the case of degenerate eigenvalues we assume that the eigenvectors have
been properly orthogonalised.

The linearity of the trace gives
\begin{equation*}
  S(\rho \| \sigma)
  = \tr\{ \rho \log \rho \} - \tr\{ \rho \log \sigma \}
\end{equation*}
where the first term can be identified as $-S(\rho)$, where
$S(\cdot)$ is the von Neumann entropy. This depends only on the
eigenvalues of $\rho$:
\begin{equation*}
  \tr\{\rho \log \rho \} = \sum_i r_i \log r_i
  .
\end{equation*}

The second term is more complicated as the eigenvectors of the density
operators in general are not the same. We decompose the density operators
using the spectral decomposition.
\begin{align*}
  \rho \log \sigma
  &= \Big( \sum_i r_i \ket{r_i}\bra{r_i} \Big)
  \log \Big( \sum_j s_j \ket{s_j}\bra{s_j} \Big)\\
  &= \Big( \sum_i r_i \ket{r_i}\bra{r_i} \Big)
  \Big( \sum_j \log (s_j) \ket{s_j}\bra{s_j} \Big)\\
  &= \sum_{ij} r_i \log(s_j) \braket{r_i|s_j} \ket{r_i}\bra{s_j}\\
  \intertext{
    Then we decompose $\ket{r_i}$ in the eigendirections of $\sigma$,
    $\ket{r_i} = \sum_k \braket{s_k|r_i} \ket{s_k}$.
  }
  \rho \log \sigma
  &= \sum_{ijk} r_i \log(s_j) \braket{r_i|s_j} \braket{s_k|r_i}
  \ket{s_k}\bra{s_j}
\end{align*}

Now we can take the trace of this to get the second term
\begin{equation*}
  \tr\{ \rho \log \sigma \}
  =
  \sum_{ij} r_i \log s_j \left|\braket{r_i|s_j}\right|^2
\end{equation*}

We can now put the two terms together to get the expression we wanted;
\begin{equation}
  \label{eq:rel_ent_expr}
  S(\rho \| \sigma)
  =
  \sum_i r_i \bigg\{
  \log r_i - \sum_j \log s_j \left| \braket{r_i|s_j}\right|^2
  \bigg\}
  .
\end{equation}

This may also be written as the single sum
\begin{equation}
  \label{eq:rel_ent_expr_single}
  S(\rho \| \sigma)
  =
  \sum_i \bigg\{
  r_i \log r_i - \bra{s_i}\rho\ket{s_i} \log s_i
  \bigg\}
  .
\end{equation}

In the case when the eigenvectors for the two density matrices are
the same (i.e.~$\rho$ and $\sigma$ commute), we can arrange
the eigenvectors so that $\braket{r_i|s_j} = \delta_{ij}$.
\eqref{eq:rel_ent_expr} then reduces to 
\begin{equation}
  \label{eq:rel_ent_expr_special}
  S(\rho \| \sigma)
  =
  \sum_i r_i \big(
  \log r_i - \log s_i
  \big)
  .
\end{equation}
\cleardoublepage

\bibliography{report}
\bibliographystyle{hep}
\cleardoublepage

% Index
\printindex

\end{document}